\documentclass[aps,pra,showpacs,amsmath,amssymb,amsfonts,tightenlines,twocolumn,lengthcheck]{revtex4-1}
\usepackage{diagbox}
\usepackage{multirow}
\usepackage{amsmath}
\usepackage{amsfonts}
\usepackage{graphicx}
\usepackage{epsfig}
\usepackage{color}
\usepackage{txfonts}
\usepackage[colorlinks,citecolor=blue]{hyperref}

\begin{document}
\title{Non-Markovian disentanglement dynamics in double-giant-atom waveguide-QED systems}
\author{Xian-Li Yin}
\affiliation{Key Laboratory of Low-Dimensional Quantum Structures and Quantum Control of Ministry of Education, Key Laboratory for Matter Microstructure and Function of Hunan Province, Department of Physics and Synergetic Innovation Center for Quantum Effects and Applications, Hunan Normal University, Changsha 410081, China}
\author{Wen-Bin Luo}
\affiliation{Key Laboratory of Low-Dimensional Quantum Structures and Quantum Control of Ministry of Education, Key Laboratory for Matter Microstructure and Function of Hunan Province, Department of Physics and Synergetic Innovation Center for Quantum Effects and Applications, Hunan Normal University, Changsha 410081, China}
\author{Jie-Qiao Liao}
\email{Corresponding author: jqliao@hunnu.edu.cn}
\affiliation{Key Laboratory of Low-Dimensional Quantum Structures and Quantum Control of Ministry of Education, Key Laboratory for Matter Microstructure and Function of Hunan Province, Department of Physics and Synergetic Innovation Center for Quantum Effects and Applications, Hunan Normal University, Changsha 410081, China}

\begin{abstract}
We study the disentanglement dynamics of two giant atoms coupled to a common one-dimensional waveguide. We focus on the non-Markovian retarded effect in the disentanglement of the two giant atoms by taking the photon transmission time into account. By solving the time-delayed equations of motion for the probability amplitudes, we obtain the evolution of the entanglement of the two giant atoms, which are initially in the maximally entangled states in the single-excitation space. It is found that the retardation-induced non-Markovianity leads to non-exponential decay and revivals of entanglement.  Concretely, we consider separate-, braided-, and nested-coupling configurations, and find that the disentanglement dynamics in these configurations exhibits different features. We demonstrate that the steady-state entanglement depends on the time delay under certain conditions in these three coupling configurations. We also study the dependence of the disentanglement of the two giant atoms on both the detuning effect and the initial-state phase effect. In addition, we consider the disentanglement dynamics of the two giant atoms, which are initially in the state superposed by zero-excitation and two-excitation components. This work will pave the way for the generation of stationary entanglement between two giant atoms, which may have potential applications in the construction of large-scale quantum networks based on the giant-atom waveguide-QED systems.
\end{abstract}

\date{\today}
\maketitle
\section{Introduction}
Quantum entanglement~\cite{Schroedigner35,Einstein35,Horodecki09}, as an important physical resource for quantum technology, plays a critical role in both the fundamental quantum theory and quantum information science.  A lot of theoretical and experimental schemes were proposed to generate quantum entanglement in various quantum systems, such as atom-cavity systems~\cite{Haroche2001,Kuzmich2005,Weinfurter2006,Rempe2008}, trapped ion systems~\cite{Wineland2003,Monroe2004}, quantum dots~\cite{Gao2012,Lodahl15}, and superconduction qubits~\cite{Nori2011,Nori2013,Gu2017}. Waveguide quantum electrodynamics (QED) systems, as promising candidates for generating entanglement between distant atoms, attracted much attention in recent years~\cite{Gu2017,Roy2017,Sheremet2021}. Many interesting phenomena were found in waveguide-QED systems, including the few-photon transport~\cite{Fan05OL,Fan05PRL,Fan07PRA,Liao09,Liao10a,Liao10b,Liao13,Hu18,Trivedi18,Stolyarov19,Joanesarson20}, the spontaneous entanglement generation~\cite{Fleischhauer2010,Vidal2011,MartinCano2011,Porras2013,Ballestero2014}, the creation of the super- and subradiant states~\cite{Brewer96,Blais2013,Zanner22}, and the long-distance entanglement between remote atoms~\cite{Baranger2013,Facchi16}. Therefore, the waveguides can be used as excellent platforms for constructing large-scale quantum network~\cite{Kimble08} and for implementing quantum information processing~\cite{Baranger13,Tudela2016}.

In most previous schemes for generating entanglement in waveguide-QED systems, the atoms were typically considered as point-like objects, and hence the dipole approximation was usually used~\cite{Wall2008}. In recent years, giant atoms as a new research field, gained increasing attention from the peers of quantum optics~\cite{Kockum2020Rev}. In general, the giant atoms were coupled to a waveguide at multiple points. So far, many theoretical giant-atom schemes were proposed~\cite{Kockum14PRA,Guo17PRA,Kockum18,Cirac19,Johansso19,Hammerer19,Zoller20,Guo20prr,Guo20pra,Wang20,Ciccarello1,Ciccarello2,Longhi20,Kockum20,Wang21pra,WangX21,Zhu21,Wang21,Du21pra1,Du21pra2,Jia21,Vega21,Kockum22pra,Du22A,Liao22,WangX22,Du22,Santos22,Soro2022,Lim22,Yin22}, with many interesting findings including frequency-dependent Lamb shifts and relaxation~\cite{Kockum14PRA}, non-exponential atomic decay~\cite{Guo17PRA,Guo20pra,Longhi20,Delsing19,Du22A}, decoherence-free coupling between two braided giant atoms~\cite{Kockum18,Ciccarello1,Ciccarello2,Kockum22pra,Soro2022}, the formation of  bound states~\cite{Guo20prr,Wang20,WangX21,Wang21,Vega21,Lim22}, and the single-photon scattering~\cite{Wang20,Kockum20,Zhu21,Du21pra2,Jia21,Liao22}. Owing to the advancement of modern quantum technology, the giant atoms were realized in various experimental platforms~\cite{Delsing14,Aref16,Leek17,Cleland19,Delsing19,Cleland20,Delsing20,Oliver20,Wilson21}, via coupling the superconducting qubits to the surface acoustic waves (SAWs) or microwave waveguides.

It was reported in previous investigations that the non-negligible non-Markovian retarded effect can modify the dynamics of the system, such as the spontaneous emission of a single atom in front of a mirror~\cite{Tufarelli13,Tufarelli14,Zoller17}, the giant atom decay in coupled waveguide arrays~\cite{Longhi20,Lim22}, the collective radiation from two separate small atoms~\cite{Solano20L,Solano20A} or giant atoms~\cite{lv22arx}, and the single-photon nonreciprocal excitation transfer between emitters in waveguide-QED systems~\cite{Du21pra1}. Meanwhile, the disentanglement dynamics with non-Markovian effect was also studied in a two-qubit system~\cite{Compagno07,Zheng08,Plastina08,Mazzola09,Moreno13}. However, for the giant atoms coupled to a waveguide at multiple coupling points, how the striking interference and the retardation-induced non-Markovianity jointly affect the disentanglement dynamics of two atoms remains an unknown and interesting topic. Note that the disentanglement dynamics of two small atoms coupled to various environments was studied and interesting effects were found in these systems~\cite{Yu04,Yu06,Roszak06,Sun07,Zubairy07,Yu09}, such as entanglement sudden death and entanglement collapse and revival.

In this paper, we study the disentanglement dynamics of two giant atoms coupled to a common waveguide. Here, the two giant atoms are initially in maximally entangled states. We show that the non-negligible non-Markovian effect can give rise to non-exponential decay and revivals of entanglement in the double-giant-atom waveguide-QED system, where each giant atom interacts with the waveguide at two separate coupling points. The different arrangement of the coupling points gives three different coupling configurations: the separate, braided, and nested couplings~\cite{Kockum18}. We find that the changes of the phase shift, the time delay, the atomic initial state, and the coupling configurations can lead to the transition from the exponential decay or non-exponential decay of the entanglement to the steady-steady entanglement. By restoring to the final-value theorem~\cite{Gluskin03}, we obtain the steady-state entanglement between the two giant atoms, which depends on the time delay and the different coupling configurations. In addition, by introducing the Dicke symmetric and antisymmetric states in the cases of separate and braided giant atoms, we find that the equations of motion for the amplitudes of the Dicke states are decoupled. Particularly, for the two atoms initially in the symmetric state, the disentanglement dynamics is governed by the same equation for the amplitude of the symmetric state. The effect of the frequency detuning of the giant atoms on the disentanglement dynamics is also analyzed. In the cases of the braided and nested coupling configurations, the disentanglement dynamics between the two giant atoms can exhibit different features from small atoms. For a general entangled state with a phase, we obtain the dependence of the steady-state entanglement on the phase. Finally, we show that the steady-state entanglement can be obtained by numerically solving the time-delayed quantum master equation when the two giant atoms are initially in the state superposed by zero-excitation and two-excitation components.

The rest of this paper is organized as follows. In Sec.~\ref{Physical model and Eqs}, we introduce the physical system for two giant atoms coupled to a common waveguide and present the Hamiltonian. In Sec.~\ref{Dynamics of disEn A}, we study the influence of the phase shift, the time delay, and the coupling configurations on the disentanglement dynamics of two giant atoms in the single-excitation subspace. In Sec.~\ref{Effect of the detuning}, we present some analysis on the disentanglement dynamics of two giant atoms with different transition frequencies. In Sec.~\ref{Effect of the phase}, we consider the case, where the two giant atoms are initially in a general entangled state with a phase in the single-excitation subspace. In Sec.~\ref{Double-extication state}, we study the disentanglement dynamics of the giant atoms starting in the state with two-atom ground-state and excited-state components. Finally, we present a brief discussion and conclusion in Sec.~\ref{Discussion and conclusion}.

\begin{figure}[tbp]
\center\includegraphics[width=0.48\textwidth]{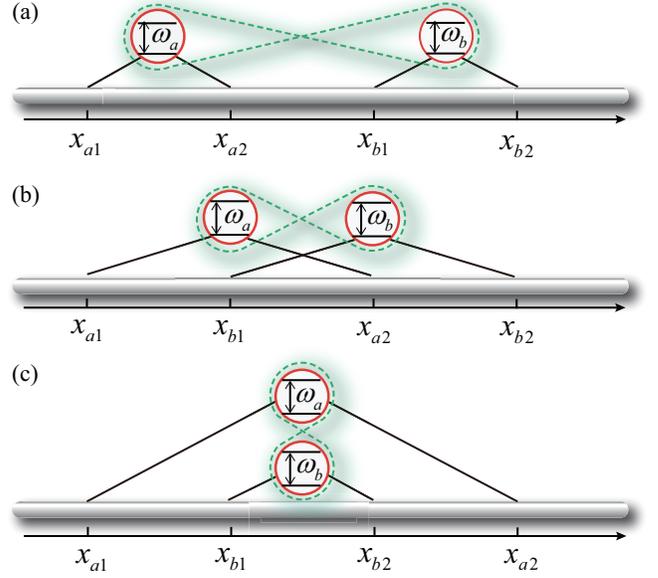}
\caption{Schematic of the two-giant-atom waveguide-QED systems. The two giant atoms, labeled $a$ and $b$, are coupled to a  waveguide through the (a) separate-,  (b) braided-,  and (c) nested-coupling configurations. In all panels, the two giant atoms are initially prepared in various entangled states.}
\label{modelv1}
\end{figure}

\section{System and Hamiltonians}\label{Physical model and Eqs}
We start by considering a two-giant-atom waveguide-QED system, in which each giant atom couples to a common waveguide through two separate coupling points. According to the different coupling arrangement of the two giant atoms with the waveguide, there exist three coupling configurations~\cite{Kockum18}:  the separate, braided, and nested couplings, as shown in Figs.~\ref{modelv1}(a) to \ref{modelv1}(c). The coordinates of the coupling points are denoted by $x_{jn}$, with $j=a,b$ marking the giant atom and $n=1,2$ representing the coupling points. In the rotating-wave approximation, the Hamiltonian of the system reads ($\hbar=1$)
\begin{eqnarray}
\label{Hamiltonian in real space}
\hat{H} &=&-i\upsilon _{g}\int dx\hat{c}_{R}^{\dagger }(x)\frac{\partial }{\partial x}\hat{c}_{R}(x)  \nonumber \\
&&+i\upsilon _{g}\int dx\hat{c}_{L}^{\dagger }(x)\frac{\partial }{\partial x}
\hat{c}_{L}(x)+\sum_{j=a,b}\omega _{j}\hat{\sigma}_{j}^{+}\hat{\sigma}_{j}^{-}  \nonumber \\
&&+\sum_{o=L,R}\sum_{j=a,b}\sum_{n=1,2}\int dx g_{jn}\delta (x-x_{jn})\left[ \hat{c}_{o}^{\dagger }(x)\hat{\sigma}_{j}^{-}+\text{H.c.}\right] ,\nonumber \\
\end{eqnarray}
where $\upsilon _{g}$ is the group velocity of the photons propagating in the waveguides.  The operators $\hat{c}_{R}(x)$ $[\hat{c}_{R}^{\dagger }(x)]$ and $\hat{c}_{L}(x)$ $[\hat{c}_{L}^{\dagger }(x)]$ are the field operators describing the annihilation (creation) of a right- and left-propagating photon at position $x$ in the waveguide, respectively. The $\hat{\sigma}_{j}^{+}=|e\rangle _{jj}\langle g|$ and $\hat{\sigma}_{j}^{-}=|g\rangle_{jj}\langle e|$ are the raising and lowering operators of the giant atom $j=a$, $b$, respectively, and $\omega _{j}$ is the transition frequency between the excited state $|e\rangle _{j}$ and ground state $|g\rangle _{j}$. The $\delta(x)$ is the Dirac delta function and $g_{jn}$ is the coupling strength related to the coupling point $x_{jn}$.

To investigate the dynamics of the two giant atoms, we present the expression of the Hamiltonian in the momentum space. According to the method used in Ref.~\cite{Fan07PRA}, we introduce the Fourier transformation
\begin{equation}
\hat{c}_{R}(x)=\sum_{k_{R}}\hat{c}_{k_{R}}e^{ik_{R}x},\hspace{0.4cm}\hat{c}_{L}(x)=\sum_{k_{L}}\hat{c}_{k_{L}}e^{ik_{L}x}\label{Fouier Transform}
\end{equation}
for the operators $\hat{c}_{R}(x)$ and  $\hat{c}_{L}(x)$. Note that we use the summation of the wave vector $k_{R}$ and $k_{L}$ in Eq.~(\ref{Fouier Transform}) for the convenience of our analysis. In later calculations, we will replace the summation of $k_{R}$ and $k_{L}$ with their integral. The operator $\hat{c}_{k_{R}}$ $(\hat{c}_{k_{L}})$ in Eq.~(\ref{Fouier Transform}) denotes the annihilation operator for the right (left)-propagating photon with wave vector $k_{R}(>0)$ $[k_{L}(<0)]$ and frequency $\omega_{k_{R}}=\upsilon _{g}k_{R}$ $(\omega _{k_{L}}=-\upsilon _{g}k_{L})$. If we are interested in a narrow bandwidth in the vicinity of $\omega _{j}$, the range of $k_{R}$ and $k_{L}$ can be extended to $( -\infty ,\infty)$, then the Hamiltonian in Eq.~(\ref{Hamiltonian in real space}) can be written as
\begin{eqnarray}
\hat{H} &=&\sum_{j=a,b}\omega _{j}\hat{\sigma}_{j}^{+}\hat{\sigma}_{j}^{-}+\sum_{k}\omega _{k}\hat{c}_{k}^{\dagger }\hat{c}_{k}  \notag \\
&&+\sum_{k}\sum_{j=a,b}\sum_{n=1,2}(g\hat{c}_{k}\hat{\sigma}_{j}^{+}e^{ikx_{jn}}+\text{H.c.}).\label{Model1-Hamiltonian}
\end{eqnarray}
For simplicity, we assume that the coupling strengths at each coupling point are equal to $g$ in Eq.~(\ref{Model1-Hamiltonian}).

\section{Disentanglement dynamics of the two giant atoms in the single-excitation subspace}\label{Dynamics of disEn A}
In this section, we investigate the disentanglement dynamics of two giant atoms coupled to a common waveguide. In particular, we consider three different coupling configurations of the two giant atoms interacting with the waveguide. To this end, we first derive the equations of motion for the probability amplitudes of the giant atoms.

\subsection{Equations of motion for the probability amplitudes of the giant atoms}\label{Eqns of motion}
Since the total excitation number operator $\hat{N}=\sum_{j=a,b}\hat{\sigma}_{j}^{+}\hat{\sigma}_{j}^{-}+\sum_{k}\hat{c}_{k}^{\dagger }\hat{c}_{k}$ is a conserved quantity, then a general state in the single-excitation subspace of the system can be expressed as
\begin{equation}
\left\vert \Psi(t) \right\rangle =\sum_{j=a,b}c_{j}(t) e^{-i\omega _{j}t}\hat{\sigma}_{j}^{+}\left\vert G\right\rangle
+\sum_{k}u_{k}\left( t\right) e^{-i\omega _{k}t}\hat{c}_{k}^{\dagger }\left\vert G\right\rangle,\label{Wavefunction-model1}
\end{equation}
where $|G\rangle$ represents the state in which the waveguide is in a vacuum state and the giant atoms are in their ground state. The $c_{j}(t)$ is the probability amplitude of the atom $j$, and $u_{k}(t)$ denotes the single-photon probability amplitude of the mode $\hat{c}_{k}$, which satisfy the normalized condition $\sum_{j=a,b}|c_{j}(t) |^{2}+\sum_{k}\left\vert u_{k}(t)\right\vert ^{2}=1$. Based on the Schr\"{o}dinger equation $i\partial \left\vert \Psi \left( t\right) \right\rangle /\partial t=\hat{H}\left\vert \Psi \left( t\right) \right\rangle$, we obtain the equations of motion for these probability amplitudes,
\begin{eqnarray}
\dot{c}_{a}(t)&=&-i\sum_{k}gu_{k}(t)(e^{ikx_{a1}}+e^{ikx_{a2}}) e^{-i\left( \omega _{k}-\omega _{a}\right)t},  \notag \\
\dot{c}_{b}(t)&=&-i\sum_{k}gu_{k}(t) (e^{ikx_{b1}}+e^{ikx_{b2}}) e^{-i\left( \omega _{k}-\omega _{b}\right)t},  \notag \\
\dot{u}_{k}(t) &=&-i\sum_{j=a,b}gc_{j}(t)( e^{-ikx_{j1}}+e^{-ikx_{j2}}) e^{i\left( \omega_{k}-\omega _{j}\right) t}.\label{coupledEq}
\end{eqnarray}

The formal solution of $u_{k}(t)$ can be obtained as
\begin{equation}
u_{k}(t) =-i\sum_{j=a,b}\int_{0}^{t}gc_{j}(t^{\prime })( e^{-ikx_{j1}}+e^{-ikx_{j2}}) e^{i(\omega _{k}-\omega_{j}) t^{\prime }}dt^{\prime },\label{fieldEq}
\end{equation}
 where we assume that $u_{k}(0)=0$, which means that the waveguide is initially in a vacuum state. We now introduce the relations
 \begin{eqnarray}
c_{a}(t) &=&\tilde{c}_{a}(t) e^{i\delta t},  \nonumber \\
c_{b}(t)&=&\tilde{c}_{b}(t)e^{-i\delta t},\label{tilde_ca_cb}
\end{eqnarray}
where  $\delta=(\omega_{a}-\omega_{b})/2$ is defined as the frequency detuning and meanwhile we introduce the mean frequency $\omega_{0}=(\omega_{a}+\omega_{b})/2$. Using the Wigner-Weisskopf approximation~\cite{Scullybook} and assuming $\omega_{k}\approx \omega _{0}+\nu=\omega_{0}+(k-k_{0})\upsilon_{g}$~\cite{Fan05PRL}, with $k_{0}$ ($\upsilon_{g}$) being the wave vector (group velocity) of the field at frequency $\omega_{0}$, one can obtain the time-delayed differential equations of the probability amplitudes as
\begin{subequations}
\begin{align}
\dot{\tilde{c}}_{a}(t)=& -\gamma \tilde{c}_{a}(t) -i\delta \tilde{c}_{a}(t)-\gamma e^{i\theta _{0}^{(a)}}\tilde{c}_{a}(t-t_{d}^{(a)})\Theta (t-t_{d}^{(a)})\nonumber \\
& -\frac{\gamma }{2}\sum_{n,n^{\prime }=1,2}e^{i\theta _{0}^{(an,bn^{\prime})}}\tilde{c}_{b}(t-t_{d}^{(an,bn^{\prime })})\Theta(t-t_{d}^{(an,bn^{\prime })}),  \label{CoupleEqforTGAa} \\
\dot{\tilde{c}}_{b}(t)=& -\gamma\tilde{c}_{b}(t) +i\delta\tilde{c}_{b}(t)-\gamma e^{i\theta _{0}^{(b)}}\tilde{c}_{b}(t-t_{d}^{(b)})\Theta (t-t_{d}^{(b)})\nonumber \\
& -\frac{\gamma }{2}\sum_{n,n^{\prime }=1,2}e^{i\theta _{0}^{(an,bn^{\prime})}}\tilde{c}_{a}(t-t_{d}^{(an,bn^{\prime })})\Theta(t-t_{d}^{(an,bn^{\prime })}),  \label{CoupleEqforTGAb}
\end{align}
\end{subequations}
where $\Theta(t)$ is the Heaviside step function and $\gamma =4\pi g^{2}/\upsilon _{g}$ is the atomic spontaneous emission rate. Note that in the derivation of  Eqs.~(\ref{CoupleEqforTGAa}) and~(\ref{CoupleEqforTGAb}), we consider the case of the weak-coupling regime, which is a common strategy that was used in many previous studies of small atoms~\cite{Tufarelli13,Tufarelli14,Zoller17} and giant atoms~\cite{Guo20prr,Du21pra1,Du22A}. We would like to point out that the ultrastrong coupling in giant atoms may be an interesting topic for future research~\cite{Noachtar22,Zhang22,Zueco22}. In addition,
in Eqs.~(\ref{CoupleEqforTGAa}) and~(\ref{CoupleEqforTGAb}) we neglect the internal dissipation of the giant atoms, by assuming it is much weaker than the spontaneous emission rate, which is a standard approximation used in most of the current theoretical works~\cite{Kockum14PRA,Guo17PRA,Kockum18,Johansso19,Guo20prr,Guo20pra,Wang20,Ciccarello1,Ciccarello2,Longhi20,Kockum20,WangX21,Zhu21,Wang21,Du21pra1,Du21pra2,Jia21,Vega21,Kockum22pra,Liao22,WangX22}. Meanwhile, this approximation is also reasonable in the platform of the superconducting qubits, since the lifetime of the excited state is much longer than the time scale for the coherence processes of the system. In Eqs.~(\ref{CoupleEqforTGAa}) and~(\ref{CoupleEqforTGAb}), we introduce the accumulated phase shift $\theta _{0}^{(j)} =k_{0}|x_{j1}-x_{j2}|$ ($\theta _{0}^{\left( an,bn^{\prime }\right) } =k_{0}|x_{an}-x_{bn^{\prime }}|$) and the time delay $t_{d}^{\left( j\right) }=|x_{j1}-x_{j2}| /\upsilon_{g}$ ($t_{d}^{\left( an,bn^{\prime }\right) } =\left \vert x_{an}-x_{bn^{\prime}}\right \vert /\upsilon _{g}$) of photons propagating between the inner (any two) coupling points of each giant atom (two giant atoms).

The first term at the right-hand side of Eqs.~(\ref{CoupleEqforTGAa}) and~(\ref{CoupleEqforTGAb})  corresponds to the typical spontaneous emission of a two-level atom with a damping rate $2\gamma$. The second term is caused by the frequency detuning of the two giant atoms. The third term describes the process that the emitted photon is re-absorbed by the giant atom $j$ itself at times $t\geq t_{d}^{(j)}$ due to the existence of two coupling points. The second line in Eqs.~(\ref{CoupleEqforTGAa}) and~(\ref{CoupleEqforTGAb}) indicates that the giant atom is re-excited by the other one when $t>t_{d}^{(an,bn^{\prime })}$. Due to the existence of multiple coupling points of the giant atoms, quantum interference plays an important role in the system, and more complicated re-emissions and re-absorptions of photons take place. Moreover, if the propagating times $t_{d}^{(j)}$ and $t_{d}^{(an,bn^{\prime })}$ are non-negligible compared to the lifetime of the giant atoms, the non-Markovian retarded effect should be taken into account. Then the system will exhibit some non-Markovian features different from the Markovian case.

\subsection{Reduced density matrix and concurrence of the two giant atoms}\label{Eqns of motion}
To characterize the quantum entanglement, we adopt the concurrence to quantitively measure quantum entanglement between the two giant atoms~\cite{Wootters98}. For simplicity, we assume that the distances between adjacent coupling points are equal to $d$ with corresponding phase shift $\theta_{0}=k_{0}d$. In the bases of $\left\{ \left\vert e\right\rangle _{a}\left\vert
e\right\rangle _{b},\left\vert e\right\rangle _{a}\left\vert g\right\rangle_{b},\left\vert g\right\rangle _{a}\left\vert e\right\rangle _{b},\left\vert g\right\rangle _{a}\left\vert g\right\rangle _{b}\right\} $, the reduced density matrix of the two giant atoms in state~(\ref{Wavefunction-model1}) is given by
\begin{equation}
\hat{\rho}(t) =\left(
\begin{array}{cccc}
0 & 0 & 0 & 0 \\
0 & |\tilde{c}_{a}(t) |^{2} & \tilde{c}_{a}(t) \tilde{c}_{b} ^{\ast}(t) & 0 \\
0 & \tilde{c}_{a}^{\ast}(t)\tilde{c}_{b}(t) & |\tilde{c}_{b}(t) |^{2} & 0 \\
0 & 0 & 0 & 1-|\tilde{c}_{a}(t) |^{2}-|\tilde{c}_{b}(t) |^{2}
\end{array}
\right),\label{reduced Density MT}
\end{equation}
where $\tilde{c}_{a}(t) $ and $\tilde{c}_{b}(t) $ are introduced in Eq.~(\ref{Wavefunction-model1}). For the density matrix $\hat{\rho}(t)$, the concurrence can be calculated as
\begin{equation}
C(t) =2|\tilde{c}_{a}(t) \tilde{c}_{b}^{\ast }(t)|.\label{Concurrence}
\end{equation}

The expressions of $\tilde{c}_{a}(t)$ and $\tilde{c}_{b}(t)$ can be obtained by solving Eqs.~(\ref{CoupleEqforTGAa}) and (\ref{CoupleEqforTGAb}) under the initial condition, and then the concurrence $C(t)$ can be calculated by Eq.~(\ref{Concurrence}). Further, we can study the effect of the phase shift, the time delay, and the coupling configurations on the disentanglement dynamics of the two giant atoms. In the single-excitation case, the two giant atoms can be assumed in a general entangled state $|\psi\rangle _{+}=(|e\rangle _{a}|g\rangle _{b}+ e^{i\phi }|g\rangle _{a}|e\rangle _{b})/\sqrt{2}$. However, for simplicity, we first consider the case of the phase $\phi=0$ and $\pi$, which corresponds to the symmetric and antisymmetric states $|\pm\rangle=(|e\rangle_{a}|g\rangle_{b}\pm|g\rangle_{a}|e\rangle_{b})/\sqrt{2}$, respectively. The influence of the phase $\phi$ will be discussed in detail in Sec.~\ref{Effect of the phase}. Accordingly, the equations of motion for the probability amplitudes in Eqs.~(\ref{CoupleEqforTGAa}) and~(\ref{CoupleEqforTGAa}) can be re-expressed with the variables $\alpha _{+}(t)=[\tilde{c}_{a}(t)+\tilde{c}_{b}(t)]/\sqrt{2}$ and $\alpha _{-}(t)=[\tilde{c}_{a}(t)-\tilde{c}_{b}(t)]/\sqrt{2}$, where $\alpha _{+}(t)$ and $\alpha _{-}(t)$ are the amplitudes of the symmetric and antisymmetric states, respectively. In the following, we will investigate the disentanglement dynamics of the two giant atoms in three different coupling configurations.
\begin{figure}[tbp]
\center\includegraphics[width=0.48\textwidth]{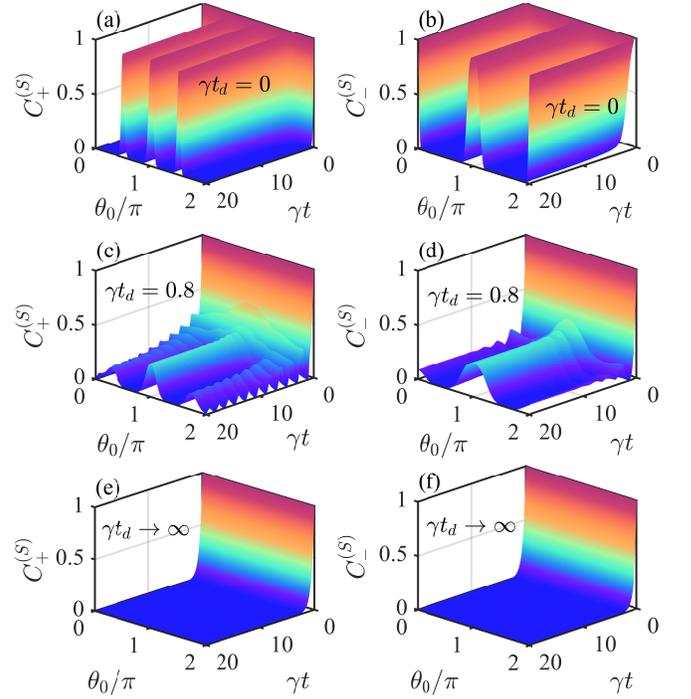}
\caption{Concurrences $C^{(S)}_{\pm}$ as functions of the evolution time $\gamma t$ and the phase shift $\theta_{0}$ at different values of the time delay $\gamma t_{d}$. The left and right columns correspond to the initial symmetric state $|+\rangle$ and antisymmetric state $|-\rangle$, respectively. The time delay $\gamma t_{d}=0$, $0.8$, and $\infty$ are set in panels (a,b), (c,d), and (e,f), respectively.}
\label{CS-vs-tandtheta-3D}
\end{figure}

\subsection{Disentanglement dynamics of the two separate giant atoms}
We first consider the case of two separate giant atoms, as shown in Fig.~\ref{modelv1}(a). Based on Eqs.~(\ref{CoupleEqforTGAa}) and~(\ref{CoupleEqforTGAb}), the equations of motion for the amplitudes of the symmetric and antisymmetric states can be obtained as
\begin{subequations}
\begin{align}
\dot{\alpha}_{+}^{(S)}(t)  &=-\gamma \alpha _{+}^{(S)}(t)-i\delta \alpha _{-}^{(S)}(t)-\frac{3\gamma }{2}\Lambda _{1,+}^{(S)}-\gamma \Lambda
_{2,+}^{(S)}-\frac{\gamma }{2}\Lambda _{3,+}^{(S)},\label{S-ANS-Sp} \\
\dot{\alpha}_{-}^{(S)}(t)  &=-\gamma \alpha _{-}^{(S)}(t)-i\delta \alpha _{+}^{(S)}(t) -\frac{\gamma }{2}\Lambda _{1,-}^{(S)}+\gamma \Lambda _{2,-}^{(S)}+
\frac{\gamma }{2}\Lambda _{3,-}^{(S)}\label{S-ANS-Sm},
\end{align}
\end{subequations}
where $\Lambda _{l,\pm}^{(S)}=e^{il\theta _{0}}\alpha_{\pm}^{(S)}( t-ld/\upsilon_{g}) \Theta( t-ld/\upsilon _{g})$ with $l=1,2$, and $3$. Here, the superscript $S$ represents the separate-coupling configuration, while the subscripts $l$ and $\pm$ are used to denote the time delay $lt_{d}$ and the atomic initial state, respectively. For simplicity, we start by considering the case where the two giant atoms have the same transition frequency, i.e., $\delta=\omega_{a}-\omega_{b}=0$. Then,  it can be seen from Eqs.~(\ref{S-ANS-Sp}) and~(\ref{S-ANS-Sm}) that the equations of motion for the amplitudes $\alpha_{\pm}^{(S)}(t)$ are decoupled from each other.

Based on Eqs.~(\ref{S-ANS-Sp}) and~(\ref{S-ANS-Sm}), the corresponding Laplace transforms of $\alpha_{\pm}^{(S)}(t)$ in the case of $\delta=0$ can be obtained as
\begin{equation}
\tilde{\alpha}_{\pm}^{(S)}(s)=\frac{\alpha _{\pm}^{(S)}(0)}{s+\gamma Y_{\pm}^{(S)}},\label{Laplace-S-alpha-pm}
\end{equation}
with
\begin{eqnarray}
Y_{+}^{(S)} &=&1+\frac{3}{2}e^{\theta}+e^{2\theta}+\frac{1}{2}e^{3\theta},\nonumber \\
Y_{-}^{(S)} &=&1+\frac{1}{2}e^{\theta}-e^{2\theta}-\frac{1}{2}e^{3\theta}.\label{YS}
\end{eqnarray}
In Eq.~(\ref{YS}), we introduce the phase $\theta=i\theta_{0}-st_{d}$.

To see the effect of the phase shift $\theta_{0}$ and the initial state of the two atoms on the disentanglement dynamics, we plot in Fig.~\ref{CS-vs-tandtheta-3D} the concurrences $C^{(S)}_{\pm}$ as functions of the evolution time $\gamma t$ and $\theta_{0}$ when $\gamma t_{d}$ takes different values. The left and right columns in Fig.~\ref{CS-vs-tandtheta-3D} correspond to the symmetric [$\alpha_{+}^{(S)}(0)=1$] and antisymmetric [$\alpha_{-}^{(S)}(0)=1$] initial states of the atoms, respectively.  For the two states, the concurrences are given by $C_{+}^{(S)}(t)=|\alpha^{(S)}_{+}(t)|^{2}$ and $C_{-}^{(S)}(t)=|\alpha^{(S)}_{-}(t)|^{2}$, respectively. It can be found from Figs.~\ref{CS-vs-tandtheta-3D}(a) and~\ref{CS-vs-tandtheta-3D}(b) that, when the time delay $\gamma t_{d}=0$, the dynamics of the concurrences $C^{(S)}_{\pm}$ are jointly determined by the phase shift and the initial condition. In Fig.~\ref{CS-vs-tandtheta-3D}(a), the concurrence remains its initial value, i.e.,  $C^{(S)}_{+}(t)=|\alpha_{+}^{(S)}(0)|^{2}=1$ as time goes at both $\theta_{0}=(m+1/2)\pi$ and $(2m+1)\pi$ for an integer $m$. When the two atoms are initially in the state $|-\rangle$, the concurrence keeps unchanged [$C^{(S)}_{-}(t)=|\alpha_{-}^{(S)}(0)|^{2}=1$] at $\theta_{0}=m\pi$. For other values of $\theta_{0}$, both the concurrences $C^{(S)}_{\pm}$ exhibit exponentially decays with time. To explain this phenomenon, we substitute $lt_{d}\rightarrow 0$ into Eqs.~(\ref{S-ANS-Sp}) and~(\ref{S-ANS-Sm}) and obtain
\begin{subequations}
\begin{align}
\dot{\alpha}_{+}^{(S)}(t)&=-\gamma \left( 1+\frac{3}{2}e^{i\theta _{0}}+e^{2i\theta _{0}}+\frac{1}{2}e^{3i\theta _{0}}\right)\alpha _{+}^{(S)}(t) , \label{alpha-a-td0-S}\\
\dot{\alpha}_{-}^{(S)}(t) &=-\gamma \left(1+\frac{1}{2}e^{i\theta _{0}}-e^{2i\theta _{0}}-\frac{1}{2}e^{3i\theta _{0}}\right)\alpha _{-}^{(S)}(t).\label{alpha-b-td0-S}
\end{align}
\end{subequations}
\begin{figure}[tbp]
\center\includegraphics[width=0.48\textwidth]{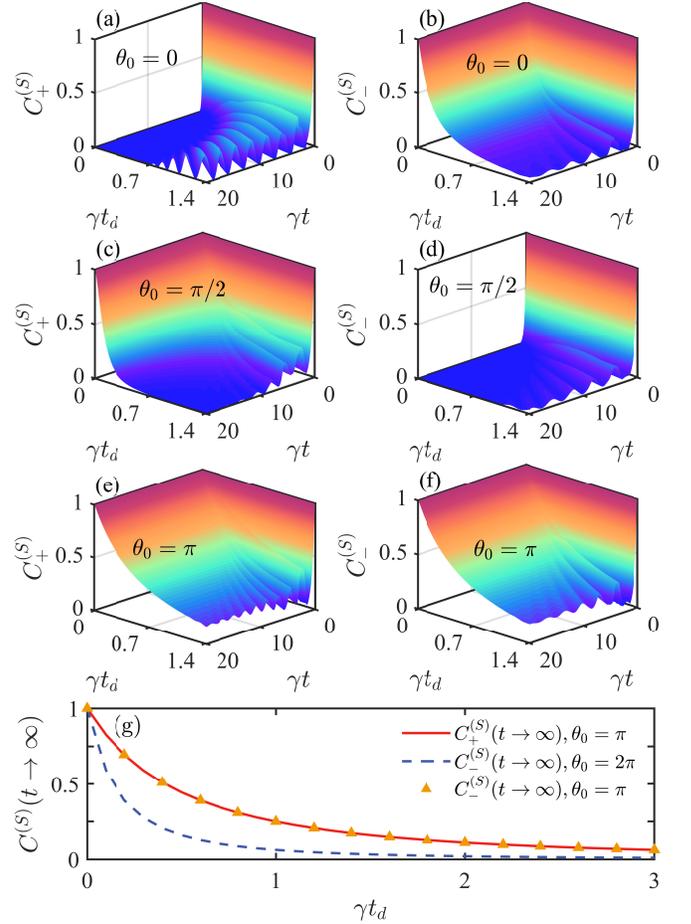}
\caption{(a)--(f) Concurrences $C^{(S)}_{\pm}$ as functions of the evolution time $\gamma t$ and the time delay $\gamma t_{d}$ at different values of $\theta_{0}$. The left and right columns correspond to the states $|+\rangle$ and $|-\rangle$, respectively. The parameters $\theta_{0}=0$, $\pi/2$, and $\pi$ are set for panels (a,b), (c,d), and (e,f) respectively. (g) The steady-state concurrences $C^{(S)}_{\pm}(t\rightarrow\infty)$ as functions of the time delay $\gamma t_{d}$ when $\theta_{0}$ takes different values.}
\label{CS-vs-tandtd-3D}
\end{figure}

The effective decay rates in Eqs.~(\ref{alpha-a-td0-S}) and~(\ref{alpha-a-td0-S}) depend on $\theta_{0}$, and hence the amplitudes will be modulated by quantum interference. Substituting $\theta_{0}=2m\pi$, $(2m+1)\pi$, and $(m+1/2)\pi$ into Eqs.~(\ref{alpha-a-td0-S}) and~(\ref{alpha-b-td0-S}), we obtain $\dot{\alpha}_{+}^{(S)}(t)=-4\gamma\alpha_{+}^{(S)}(t)$ and $\dot{\alpha}_{-}^{(S)}(t)=0$ for $\theta_{0}=2m\pi$, $\dot{\alpha}_{+}^{(S)}(t)=\dot{\alpha}_{-}^{(S)}(t)=0$ for $\theta_{0}=(2m+1)\pi$, $\dot{\alpha}_{+}^{(S)}(t)=-i\gamma\alpha_{+}^{(S)}(t)$ and $\dot{\alpha}_{-}^{(S)}(t)=-(2+i)\gamma\alpha_{-}^{(S)}(t)$ for $\theta_{0}=(2m+1/2)\pi$, and $\dot{\alpha}_{+}^{(S)}(t)=i\gamma\alpha_{+}^{(S)}(t)$ and $\dot{\alpha}_{-}^{(S)}(t)=-(2- i)\gamma\alpha_{-}^{(S)}(t)$ for $\theta_{0}=(2m+3/2)\pi$. Therefore, it can be found that, the concurrence $C_{+}^{(S)}(t)$ exhibits an exponential decay at a rate $8\gamma$, and $C_{-}^{(S)}(t)$ preserves the initial entanglement $C_{-}^{(S)}(0)=1$ when $\theta_{0}=2m\pi$. Both the $C_{+}^{(S)}(t)$ and $C_{-}^{(S)}(t)$ remain in the initial entanglement when $\theta_{0}=(2m+1)\pi$. In addition to $\theta_{0}=(2m+1)\pi$, the concurrence $C^{(S)}_{+}$ can also preserve the initial entanglement at $\theta_{0}=(m+1/2)\pi$, because the amplitude $\alpha_{+}^{(S)}(t)$ only evolves over time with either a phase $-\gamma t$ for $\theta_{0}=(2m+1/2)\pi$ or a phase $\gamma t$ for $\theta_{0}=(2m+3/2)\pi$, which will not affect the value of the entanglement. These analyses confirm the numerical simulations in Figs.~\ref{CS-vs-tandtheta-3D}(a) and~\ref{CS-vs-tandtheta-3D}(b).

As the time delay increases to $\gamma t_{d}\sim1$, as shown in Figs.~\ref{CS-vs-tandtheta-3D}(c) and~\ref{CS-vs-tandtheta-3D}(d) for $\gamma t_{d}=0.8$, we see that both the concurrences $C^{(S)}_{+}$ and $C^{(S)}_{-}$ exhibit exponential decays at a rate $2\gamma$ within $t\in(0, t_{d})$, in which the non-Markovian effect is absent. Once $t\geq t_{d}$, the time-delay-induced non-Markovian effect begins to work such that the dynamics of the concurrences $C^{(S)}_{\pm}$ are modified. For $\theta_{0}\neq(2m+1)\pi$ ($\theta_{0}\neq m\pi$), the concurrence $C^{(S)}_{+}$ ($C^{(S)}_{-}$) exhibits an oscillating decay process when $t\geq t_{d}$. The revival peaks and the oscillating amplitudes of the concurrence are different for different values of $\theta_{0}$ and atomic initial states.  However, it can be found that, for $\theta_{0}=(2m+1)\pi$ ($\theta_{0}=m\pi$), the concurrence $C^{(S)}_{+}$ ($C^{(S)}_{-}$) can reach a steady-state value after experiencing a period of oscillation when $t\geq t_{d}$. This means that by choosing $\theta_{0}=(2m+1)\pi$ ($\theta_{0}=m\pi$) corresponding to the symmetric (antisymmetric) atomic initial state, we can obtain a subradiant state~\cite{Brewer96}, which gives rise to steady-state entanglement.

To see clearly the dependence of the stationary entanglement on the atomic initial state, the phase shift, as well as the time delay, we need to know the long-time expression of these probability amplitudes, which can be obtained with the final-value theorem~\cite{Gluskin03}
\begin{equation}
\alpha_{\pm}^{(S)}(t\rightarrow \infty)=\lim_{s\rightarrow 0}[s\tilde{\alpha}_{\pm}^{(S)}(s)].\label{F-V theorem}
\end{equation}
Substituting Eq.~(\ref{Laplace-S-alpha-pm}) into Eq.~(\ref{F-V theorem}), we obtain the relations $1+\frac{3}{2}e^{i\theta_{0}}+e^{2i\theta_{0}}+\frac{1}{2}e^{3i\theta_{0}}=0$ and $1+\frac{1}{2}e^{i\theta _{0}}-e^{2i\theta _{0}}-\frac{1}{2}e^{3i\theta_{0}}=0$ for the states $|+\rangle$ and $|-\rangle$, respectively. The solutions determined by the two conditions are given by $\theta_{0}=(2m+1)\pi$ (for $|+\rangle$) and $\theta_{0}=2m\pi$ or $\theta_{0}=(2m+1)\pi$ (for $|-\rangle$), respectively. The steady-state entanglements are obtained as
\begin{subequations}
\begin{eqnarray}
C_{+}^{(S)}(t\rightarrow\infty)= & \dfrac{1}{(1+\gamma t_{d})^{2}}, & \theta_{0}=(2m+1)\pi,\label{s-s value S 1}\\
C_{-}^{(S)}(t\rightarrow\infty)= & \left\{ \begin{array}{c}
\dfrac{1}{\left(1+3\gamma t_{d}\right)^{2}},\\
\\
\dfrac{1}{\left(1+\gamma t_{d}\right)^{2}},
\end{array}\right. & \begin{split}\theta_{0} & =\ 2m\pi,\label{s-s value S 23}\\
\\
\theta_{0} & =\ (2m+1)\pi.
\end{split}
\end{eqnarray}
\end{subequations}
Equations~(\ref{s-s value S 1}) and (\ref{s-s value S 23}) indicate that the concurrences $C_{\pm}^{(S)}(t)$ between the two atoms can approach stationary values when the above conditions for the phase shift $\theta_{0}$ are satisfied. Interestingly, these stationary values only depend on the time delay $\gamma t_{d}$. When $\theta_{0}=(2m+1)\pi$, we find that the concurrences $C^{(S)}_{\pm}(t)$ are characterized by the identical stationary value $1/\left( 1+\gamma t_{d}\right) ^{2}$. However, for $\theta_{0}=2m\pi$, we observe that there is only a steady-state value $1/\left( 1+3\gamma t_{d}\right) ^{2}$ for $C^{(S)}_{-}(t)$, which confirms our numerical simulations in Figs.~\ref{CS-vs-tandtheta-3D}(a) to~\ref{CS-vs-tandtheta-3D}(d). In addition, Eq.~(\ref{s-s value S 23}) implies that when the two separate giant atoms are initially in the antisymmetric state, the steady-state entanglement can appear at $\theta_{0}=2m\pi$ and $(2m+1)\pi$, while for small atoms, it can only appear at $\theta_{0}=2m\pi$.

Figures~\ref{CS-vs-tandtheta-3D}(e) and~\ref{CS-vs-tandtheta-3D}(f) show the concurrences $C^{(S)}_{\pm}$ versus $\gamma t$ and $\theta_{0}$ when the time delay $\gamma t_{d}\rightarrow\infty$. In this situation, the photons emitted by the giant atoms in the waveguide cannot be re-absorbed by the atoms. Therefore, we observe that both $C^{(S)}_{+}$ and $C^{(S)}_{-}$ exhibit an exponential decay at a decay rate $2\gamma$ as time increases and they are independent of $\theta_{0}$.

In the following, we investigate the influence of the time delay $\gamma t_{d}$ on the concurrences $C^{(S)}_{\pm}$ at different values of the phase shift $\theta_{0}$. For $\theta_{0}=2m\pi$, as shown in Figs.~\ref{CS-vs-tandtd-3D}(a) and~\ref{CS-vs-tandtd-3D}(b), both $C^{(S)}_{+}$ and $C^{(S)}_{-}$ exhibit many revival peaks and the distance between the peaks increases with the increase of $\gamma t_{d}$, which indicates that the non-Markovian retarded effect works in this case. Nevertheless, the $C^{(S)}_{+}$ cannot preserve a steady-state value after experiencing an oscillating decay while the $C^{(S)}_{-}$ can approach to a stationary value $1/\left( 1+\gamma t_{d}\right) ^{2}$. In Figs.~\ref{CS-vs-tandtd-3D}(c) and~\ref{CS-vs-tandtd-3D}(d), we take the phase shift $\theta_{0}=\pi/2$ and find that the $C^{(S)}_{+}$ can hold a fairly large value when $\gamma t_{d}\ll 1$. As $\gamma t_{d}$ increases further to $\gamma t_{d}\sim 1$,  $C^{(S)}_{+}$ is characterized by a fast oscillating decay process. For the concurrence $C^{(S)}_{-}$, as shown in Fig.~\ref{CS-vs-tandtd-3D}(d), the $C^{(S)}_{-}$ decays fast to zero when  $\gamma t_{d}\ll 1$. As $\gamma t_{d}$ further increases to approach or even larger than 1, one can observe some revival oscillating peaks induced by the non-Markovian retarded effect.

According to Eqs.~(\ref{s-s value S 1}) and~(\ref{s-s value S 23}), we know that the concurrences $C^{(S)}_{\pm}$ share the same steady-state value when $\theta_{0}=(2m+1)\pi$. Figures~\ref{CS-vs-tandtd-3D}(e) and~\ref{CS-vs-tandtd-3D}(f) show that $C^{(S)}_{\pm}$ eventually approaches an equal value after experiencing different initial oscillations, which confirms the results derived from the final-value theorem.  Figure~\ref{CS-vs-tandtd-3D}(g) shows the steady-state concurrences $C^{(S)}_{\pm}(t\rightarrow\infty)$ as functions of the time delay $\gamma t_{d}$ when $\theta_{0}$ takes different values. It can be found that the steady-state value of $C^{(S)}_{-}(t)$ at $\theta_{0}=2m\pi$ is less than and decreases faster than that of $C^{(S)}_{-}(t)$ at $\theta_{0}=(2m+1)\pi$  for an integer $m$.

\subsection{Disentanglement dynamics of the two braided giant atoms}
For the case of two braided giant atoms [Fig.~\ref{modelv1}(b)], the equations of motion for the amplitudes of the symmetric and antisymmetric states are given by
\begin{equation}
\dot{\alpha}_{\pm }^{(B)}(t) =-\gamma \alpha_{\pm }^{(B)}(t) -i\delta \alpha _{\mp}^{(B)}(t)\mp\frac{3\gamma }{2}\Lambda _{1,\pm }^{(B)}-\gamma \Lambda _{2,\pm }^{(B)}\mp
\frac{\gamma }{2}\Lambda _{3,\pm }^{(B)},\label{S-ANS-Bpm}
\end{equation}
where $\Lambda _{l,\pm}^{(B)}=e^{il\theta _{0}}\alpha_{\pm}^{(B)}( t-ld/\upsilon_{g}) \Theta( t-ld/\upsilon _{g})$, with the superscript $B$ representing the braided-coupling configuration. By comparing Eq.~(\ref{S-ANS-Sp}) with Eq.~(\ref{S-ANS-Bpm}), we find that the equations of motion for the symmetric amplitudes $\alpha_{+}^{(S)}(t)$ and $\alpha_{+}^{(B)}(t)$ have the same form. For the case of $\delta=0$, it can be seen from Eq.~(\ref{S-ANS-Bpm}) that there is a phase difference $\pi$ between the equations of motion for the amplitudes $\alpha _{+}^{(B)}(t)$ and $\alpha _{-}^{(B)}(t)$. According to Eq.~(\ref{S-ANS-Bpm}), the Laplace transform of $\alpha_{\pm}^{(B)}(t)$ can be obtained as
\begin{equation}
\tilde{\alpha}_{\pm }^{(B)}(s) =\frac{\alpha _{\pm }^{(B)}\left(0\right) }{s+\gamma Y_{\pm }^{(B)}},\label{Laplace-B-alpha-pm}
\end{equation}
with
\begin{equation}
Y_{\pm }^{(B)}=1\pm \frac{3}{2}e^{\theta}+e^{2\theta}\pm \frac{1}{2}e^{3\theta}.
\end{equation}
\begin{figure}[tbp]
\center\includegraphics[width=0.48\textwidth]{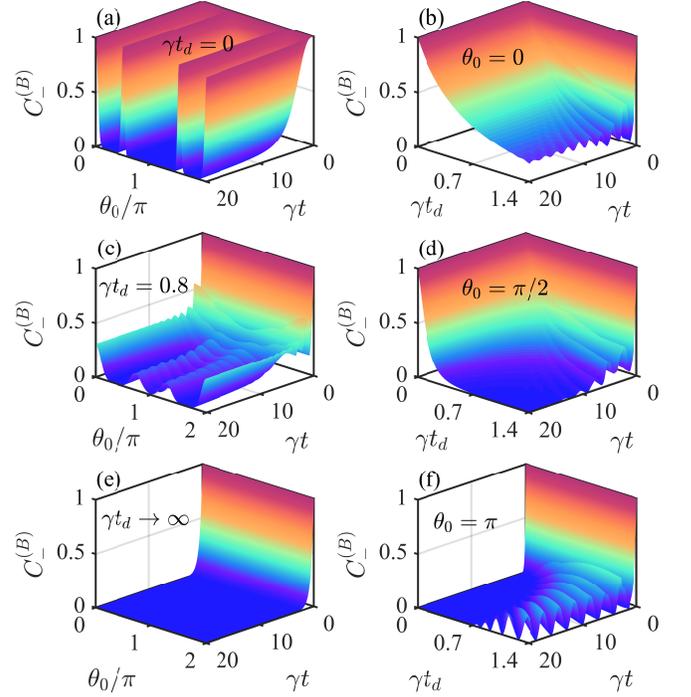}
\caption{The left column shows the concurrence $C^{(B)}_{-}$ as functions of $\gamma t$ and $\theta_{0}$ at given values of $\gamma t_{d}$. The right column shows the $C^{(B)}_{-}$ as functions of $\gamma t$ and $\gamma t_{d}$ at given values of $\theta_{0}$. The time delay $\gamma t_{d}=0$, $0.8$, and $\infty$ are set in panels (a), (c), and (e), respectively. The parameters $\theta_{0}=0$, $\pi/2$, and $\pi$ are set in panels (b), (d), and (f), respectively.}
\label{CB-tandthetatd-3D}
\end{figure}

Based on Eqs.~(\ref{F-V theorem}) and~(\ref{Laplace-B-alpha-pm}), the steady-state concurrences for the atoms initially in states $|\pm\rangle$ can be obtained by using the final-value theorem,
\begin{subequations}
\begin{align}
C_{+}^{(B)}(t &\rightarrow \infty )=\frac{1}{\left(1+\gamma t_{d}\right)^{2}},\hspace{0.25cm}\theta _{0}=(2m+1)\pi,\label{s-s value B 1} \\
C_{-}^{(B)}(t &\rightarrow \infty )=\frac{1}{\left(1+\gamma t_{d}\right)^{2}},\hspace{0.25cm}\theta _{0}=2m\pi\label{s-s value B 2}.
\end{align}
\end{subequations}
Comparing Eq.~(\ref{s-s value B 2}) with Eq.~(\ref{s-s value S 1}), we find that, the two giant atoms in both the separate and braided couplings have equal steady-state entanglement when the two atoms are initially in the symmetric state and have the phase shift $\theta_{0}=(2m+1)\pi$. This confirms our analysis concerning the equations of motion for the amplitudes $\alpha_{+}^{(S)}(t)$ and $\alpha_{+}^{(B)}(t)$. \emph{Since the amplitude $\alpha^{(B)}_{+}(t)$ has the identical time evolution with $\alpha^{(S)}_{+}(t)$, below we only focus on the antisymmetric-initial-state case in which the two atoms are initially in the antisymmetric state $|-\rangle$.}

In Figs.~\ref{CB-tandthetatd-3D}(a),~\ref{CB-tandthetatd-3D}(c), and~\ref{CB-tandthetatd-3D}(e), we show the concurrence $C_{-}^{(B)}$ versus the evolution time $\gamma t$ and the phase shift $\theta_{0}$, when the time delay is taken as $\gamma t_{d}=0$, $0.8$, and $\infty$, respectively. It can be seen from Fig.~\ref{CB-tandthetatd-3D}(a) that, when the atomic initial state is antisymmetric, the concurrence remains the initial value $C_{-}^{(B)}(t)=|\alpha_{-}^{(B)}(0)|^{2}=1$ at $\theta_{0}=2m\pi$ and $(m+1/2)\pi$. On the contrary, the concurrence exhibits an exponential decay when $\theta_{0}\neq m\pi/2$. If the time-retarded effect is negligible, i.e., $lt_{d}\rightarrow 0$, then, for the antisymmetric state, Eq.~(\ref{S-ANS-Bpm}) is reduced to
\begin{equation}
\dot{\alpha}_{-}^{(B)}(t)=-\gamma \left( 1-\frac{3}{2}e^{i\theta_{0}}+e^{2i\theta _{0}}-\frac{1}{2}e^{3i\theta _{0}}\right) \alpha_{-}^{(B)}(t).\label{alpha-td0-B}
\end{equation}
Substituting $\theta_{0}=2m\pi$ and $(m+1/2)\pi$ into Eq.~(\ref{alpha-td0-B}), we have $\dot{\alpha}_{-}^{(B)}(t)=0$ for $\theta_{0}=2m\pi$, $\dot{\alpha}_{-}^{(B)}(t)=-4\gamma\alpha_{-}^{(B)}(t)$ for $\theta_{0}=(2m+1)\pi$, $\dot{\alpha}_{-}^{(B)}(t)=i\gamma\alpha_{-}^{(B)}(t)$ for $(2m+1/2)\pi$, and $\dot{\alpha}_{-}^{(B)}(t)= -i\gamma\alpha_{-}^{(B)}(t)$ for  $(2m+3/2)\pi$. Therefore, it can be seen that the concurrence $C_{-}^{(B)}(t)$ preserves the initial entanglement for $\theta_{0}=2m\pi$ and behaves as an exponential decay at a decay rate $8\gamma$ for $\theta_{0}=(2m+1)\pi$. The concurrence can also preserve the initial value $C_{-}^{(B)}(0)=1$ when $\theta_{0}=(m+1/2)\pi$, but the amplitude $\alpha_{-}^{(B)}(t)$ evolves over time with either a phase $\gamma t$ for $\theta_{0}=(2m+1/2)\pi$ or a phase $-\gamma t$ for $\theta_{0}=(2m+3/2)\pi$.

In Fig.~\ref{CB-tandthetatd-3D}(c), we plot $C_{-}^{(B)}$ versus $\gamma t$ and $\theta_{0}$ at $\gamma t_{d}=0.8$, which shows some features different from $C_{-}^{(S)}$ due to the different coupling configurations. Since the time delay cannot be neglected, the retardation-induced non-Markovianity leads to the revival of some oscillating peaks. In particular, the $C_{-}^{(B)}$ exhibits a fast non-exponential oscillating decay process at $\theta_{0}\neq2m\pi$. Note that the non-exponential oscillating decay is weakened when $\theta_{0}=(m+1/2)\pi$. When $\theta_{0}=2m\pi$, the steady-state entanglement can also be observed for the two braided atoms, which is larger than that of the separate giant atoms [Fig.~\ref{CS-vs-tandtheta-3D}(c)]. As the time delay further increases, the concurrence $C_{-}^{(B)}$ is also characterized by an exponential decay [Fig.~\ref{CB-tandthetatd-3D}(e)], which can be explained by substituting $lt_{d}\rightarrow\infty$ into Eq.~(\ref{S-ANS-Bpm}) to obtain $\dot{\alpha}_{-}^{(B)}(t) =-\gamma \alpha _{-}^{(B)}(t)$. This indicates that the coupling configurations of the two giant atoms will not affect the dynamics of the concurrence in the infinite time delay $lt_{d}\rightarrow\infty$.

\begin{figure}[tbp]
\center\includegraphics[width=0.48\textwidth]{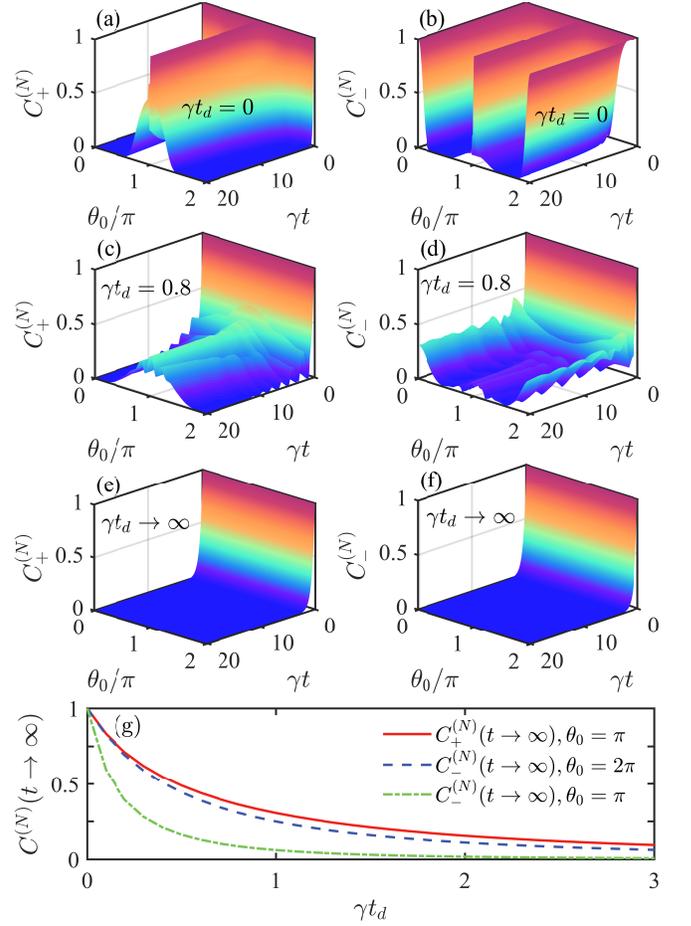}
\caption{Concurrences $C^{(N)}_{\pm}$ as functions of the evolution time $\gamma t$ and the time delay $\gamma t_{d}$ at different values of $\theta_{0}$. The left and right columns correspond to the states $|+\rangle$ and $|-\rangle$, respectively. The time delay $\gamma t_{d}=0$, $0.8$, and $\infty$ are set in panels (a,b), (c,d), and (e,f), respectively. (g) The steady-state concurrences $C^{(N)}_{\pm}(t\rightarrow\infty)$ versus the time delay $\gamma t_{d}$ when $\theta_{0}$ takes different values.}
\label{CN-vs-tandtheta-3D}
\end{figure}

Figures~\ref{CB-tandthetatd-3D}(b),~\ref{CB-tandthetatd-3D}(d), and~\ref{CB-tandthetatd-3D}(f) show the concurrence $C_{-}^{(B)}$ as a function of $\gamma t$ and $\gamma t_{d}$ when $\theta_{0}=0$, $\pi/2$, and $\pi$, respectively. Comparing these with Figs.~\ref{CS-vs-tandtd-3D}(b),~\ref{CS-vs-tandtd-3D}(d), and~\ref{CS-vs-tandtd-3D}(f), we find that the dependence of the concurrence  $C_{-}^{(B)}$ on $\gamma t$ and $\gamma t_{d}$ is the same as that of $C_{+}^{(S)}$ at $\theta_{0}=(m+1/2)\pi$. In addition, the $C_{-}^{(B)}$ at $\theta_{0}=2m\pi$ [$\theta_{0}=(2m+1)\pi$] also exhibits the same dynamics as $C_{+}^{(S)}$ at $\theta_{0}=(2m+1)\pi$ ($\theta_{0}=2m\pi$). This feature can be explained based on the existence of a phase difference $\pi$ between the equations of motion for the amplitudes $\alpha _{-}^{(B)}(t)$ and $\alpha _{+}^{(S)}(t)$, as shown in Eqs.~(\ref{S-ANS-Bpm}) and~(\ref{S-ANS-Sp}).

\begin{table*}
\caption{Concurrence between the two giant atoms for three different coupling configurations. Here we assume that the two giant atoms have the same transition frequency $\omega_{0}$ and are initially in either the symmetric or anti-symmetric state with the maximal entanglement.}
\label{table1} \centering
\begin{tabular}{|c|c|c|c|c|c|}
\hline
\multicolumn{1}{|c|}{Coupling configurations} & Time delay & \diagbox{Concurrence}{$\textrm{Phase shift \ensuremath{\theta_{0}}}$}  & $2m\pi$  & $(m+1/2)\pi$  & $(2m+1)\pi$\tabularnewline
\hline
\multirow{4}{*}{$\begin{array}{c}
\textrm{Two separate giant atoms}\end{array}$} & \multirow{2}{*}{$\gamma t_{d}>0$} & $C_{+}^{(S)}(\infty)$  & 0  & 0  & $\frac{1}{(1+\gamma t_{d})^{2}}$\tabularnewline
\cline{3-6} \cline{4-6} \cline{5-6} \cline{6-6}
 &  & $C_{-}^{(S)}(\infty)$  & $\frac{1}{(1+3\gamma t_{d})^{2}}$  & 0  & $\frac{1}{(1+\gamma t_{d})^{2}}$\tabularnewline
\cline{2-6} \cline{3-6} \cline{4-6} \cline{5-6} \cline{6-6}
 & \multirow{2}{*}{$\gamma t_{d}=0$} & $C_{+}^{(S)}(t)$  & $e^{-8\gamma t}$  & 1  & 1\tabularnewline
\cline{3-6} \cline{4-6} \cline{5-6} \cline{6-6}
 &  & $C_{-}^{(S)}(t)$  & 1  & $e^{-4\gamma t}$  & 1\tabularnewline
\hline
\multirow{4}{*}{$\begin{array}{c}
\textrm{Two braided giant atoms}\end{array}$} & \multirow{2}{*}{$\gamma t_{d}>0$} & $C_{+}^{(B)}(\infty)$  & 0  & 0  & $\frac{1}{(1+\gamma t_{d})^{2}}$\tabularnewline
\cline{3-6} \cline{4-6} \cline{5-6} \cline{6-6}
 &  & $C_{-}^{(B)}(\infty)$  & $\frac{1}{(1+\gamma t_{d})^{2}}$  & 0  & 0\tabularnewline
\cline{2-6} \cline{3-6} \cline{4-6} \cline{5-6} \cline{6-6}
 & \multirow{2}{*}{$\gamma t_{d}=0$} & $C_{+}^{(B)}(t)$  & $e^{-8\gamma t}$  & 1  & 1\tabularnewline
\cline{3-6} \cline{4-6} \cline{5-6} \cline{6-6}
 &  & $C_{-}^{(B)}(t)$  & 1  & 1  & $e^{-8\gamma t}$\tabularnewline
\hline
\multirow{4}{*}{$\begin{array}{c}
\textrm{Two nested giant atoms}\end{array}$} & \multirow{2}{*}{$\gamma t_{d}>0$} & $C_{+}^{(N)}(\infty)$  & 0  & 0  & $\frac{(1+2\gamma t_{d})(1+4\gamma t_{d})}{\left(1+4\gamma t_{d}+2\gamma^{2}t_{d}^{2}\right)^{2}}$\tabularnewline
\cline{3-6} \cline{4-6} \cline{5-6} \cline{6-6}
 &  & $C_{-}^{(N)}(\infty)$  & $\frac{1}{(1+\gamma t_{d})^{2}}$  & 0  & $\frac{(1+2\gamma t_{d})}{\left(1+4\gamma t_{d}+2\gamma^{2}t_{d}^{2}\right)^{2}}$\tabularnewline
\cline{2-6} \cline{3-6} \cline{4-6} \cline{5-6} \cline{6-6}
 & \multirow{2}{*}{$\gamma t_{d}=0$} & $C_{+}^{(N)}(t)$  & $e^{-8\gamma t}$  & $A_{+}e^{-2\gamma t}$  & 1\tabularnewline
\cline{3-6} \cline{4-6} \cline{5-6} \cline{6-6}
 &  & $C_{-}^{(N)}(t)$  & 1  & $A_{-}e^{-2\gamma t}$  & 1\tabularnewline
\hline
\end{tabular}
\end{table*}

\subsection{Disentanglement dynamics of the two nested giant atoms}
We now turn to the case of two nested giant atoms [Fig.~\ref{modelv1}(c)]. Considering the asymmetry of the equations of motion for the probability amplitudes of the two atoms, here we do not introduce the symmetric and antisymmetric states to study the disentanglement dynamics. In this case, Eqs.~(\ref{CoupleEqforTGAa}) and~(\ref{CoupleEqforTGAb}) are reduced to
\begin{subequations}
\begin{align}
\dot{c}_{a}^{(N)}(t) &=-(\gamma+i\delta) c_{a}^{(N)}(t)-\gamma \Lambda _{3,a}^{(N)}-\gamma\left( \Lambda _{1,b}^{(N)}+\Lambda_{2,b}^{(N)}\right) , \label{amplitude-ca-N} \\
\dot{c}_{b}^{(N)}(t) &=-(\gamma-i\delta) c_{b}^{(N)}(t)-\gamma \Lambda _{1,b}^{(N)}-\gamma\left( \Lambda _{1,a}^{(N)}+\Lambda_{2,a}^{(N)}\right),\label{amplitude-cb-N}
\end{align}
\end{subequations}
where $\Lambda _{l,j}^{(N)}=e^{il\theta _{0}}c_{j}^{(N)}( t-ld/\upsilon_{g}) \Theta( t-ld/\upsilon _{g})$ with $l=1,2$, and $3$ and $j=a,b$. The superscript $N$ represents the nested-coupling configuration.  According to Eqs.~(\ref{amplitude-ca-N}) and~(\ref{amplitude-cb-N}) and considering $\delta=0$, we obtain the relations
\begin{eqnarray}
\tilde{c}_{a_{+}}^{(N)}(s) &=&\frac{s+\gamma (1-e^{2\theta })}{\sqrt{2}[\gamma( s-\gamma ) e^{3\theta }-\gamma ^{2}e^{2\theta }+\gamma
(s+\gamma) e^{\theta }+(s+\gamma) ^{2}]},  \notag\\
\tilde{c}_{b_{+}}^{(N)}(s) &=&\frac{s+\gamma ( 1-e^{\theta }-e^{2\theta}+e^{3\theta }) }{\sqrt{2}[\gamma(s-\gamma) e^{3\theta
}-\gamma ^{2}e^{2\theta }+\gamma ( s+\gamma) e^{\theta }+(s+\gamma) ^{2}]},\label{Laplace-N-p}\notag\\
\end{eqnarray}
and
\begin{eqnarray}
\tilde{c}_{a_{-}}^{(N)}(s) &=&\frac{s+\gamma( 1+2e^{\theta}+e^{2\theta }) }{\sqrt{2}[\gamma (s-\gamma )e^{3\theta }-\gamma^{2}e^{2\theta }+\gamma (s+\gamma )e^{\theta }+(s+\gamma )^{2}]},  \notag \\
\tilde{c}_{b_{-}}^{(N)}(s) &=&\frac{-[s+\gamma (1+e^{\theta}+e^{2\theta }+e^{3\theta })]}{\sqrt{2}[\gamma (s-\gamma )e^{3\theta
}-\gamma ^{2}e^{2\theta }+\gamma (s+\gamma )e^{\theta }+(s+\gamma )^{2}]},\label{Laplace-N-m}\notag\\
\end{eqnarray}
for the symmetric and antisymmetric cases, respectively.

In the following, we show that the steady-state values $C_{\pm}^{(N)}(t\rightarrow\infty)$ for the nested-coupling case can also be obtained by using the final-value theorem. According to Eqs.~(\ref{Laplace-N-p}) and~(\ref{Laplace-N-m}) and utilizing $c_{j=a,b}^{(N)}(t\rightarrow \infty )=\lim_{s\rightarrow 0}[s\tilde{c}_{j}^{(N)}(s)]$, the steady-state entanglement for the two nested giant atoms are calculated as
\begin{subequations}
\begin{eqnarray}
C_{+}^{(N)}(t\!\rightarrow\!\infty)= & \dfrac{(1\!+\!2\gamma t_{d})(1\!+\!4\gamma t_{d})}{(1\!+\!4\gamma t_{d}\!+\!2\gamma^{2}t_{d}^{2})^{2}}, & \hspace{-0.15cm}\theta_{0}\!=\!(2m\!+\!1)\pi,\label{s-s value N 2}\\
C_{-}^{(N)}(t\!\rightarrow\!\infty)= & \left\{ \begin{array}{c}
\dfrac{1}{(1\!+\!\gamma t_{d})^{2}},\\
\\
\dfrac{1\!+\!2\gamma t_{d}}{(1\!+\!4\gamma t_{d}\!+\!2\gamma^{2}t_{d}^{2})^{2}},
\end{array}\right. & \hspace{-0.2cm}\begin{split}\theta_{0} & \!=2m\pi,\label{s-s value N 34}\\
\\
\theta_{0} & \!=\!(2m\!+\!1)\pi.
\end{split}
\end{eqnarray}
\end{subequations}
Note that $\theta_{0}=2m\pi$ and $\theta_{0}=(2m+1)\pi$ are the solutions of the condition $1+e^{i\theta _{0}}-e^{2i\theta _{0}}-e^{3i\theta _{0}}=0$.

Similarly, we study the dependence of $C^{(N)}_{\pm}$ on $\gamma t$ and $\theta_{0}$. In the left and right columns of Fig.~\ref{CN-vs-tandtheta-3D}, the initial state of the two atoms is assumed to be symmetric and antisymmetric, respectively. From Figs.~\ref{CN-vs-tandtheta-3D}(a) and~\ref{CN-vs-tandtheta-3D}(b), we can observe that the concurrence $C^{(N)}_{+}$ ($C^{(N)}_{-}$) preserves its initial value 1 when $\theta_{0}=(2m+1)\pi$ ($\theta_{0}=m\pi$) in the limit $\gamma t_{d}\rightarrow 0$. By taking $\gamma t_{d}=0$, Eqs.~(\ref{amplitude-ca-N}) and~(\ref{amplitude-cb-N}) become
\begin{subequations}
\begin{align}
\dot{c}_{a}^{(N)}(t) &=-\gamma(1+e^{3i\theta_{0}}) c_{a}^{(N)}(t) -\gamma(e^{i\theta_{0}}+e^{2i\theta _{0}}) c_{b}^{(N)}(t) ,\label{ca-N-td0}\\
\dot{c}_{b}^{(N)}(t)&=-\gamma(1+e^{i\theta_{0}}) c_{b}^{(N)}(t)-\gamma(e^{i\theta_{0}}+e^{2i\theta _{0}}) c_{a}^{(N)}(t).\label{cb-N-td0}
\end{align}
\end{subequations}
To remain the initial value of $C^{(N)}_{\pm}$, the phase shift $\theta_{0}$ needs to be taken as different values. If the phase shift is $\theta_{0}=(2m+1)\pi$, Eqs.~(\ref{ca-N-td0}) and~(\ref{cb-N-td0}) are reduced to $\dot{c}_{a_{\pm}}^{(N)}(t) =\dot{c}_{b_{\pm}}^{(N)}(t)=0$, which means that the unchanged $C^{(N)}_{+}$ and $C^{(N)}_{-}$ can be observed in both the symmetric and antisymmetric states, as shown in Figs.~\ref{CN-vs-tandtheta-3D}(a) and~\ref{CN-vs-tandtheta-3D}(b). However, when $\theta_{0}=2m\pi$, both  $c_{a}^{(N)}(t)$ and $c_{b}^{(N)}(t)$ are governed by the same equation
\begin{equation}
\dot{c}_{j}^{(N)}(t) =-2\gamma( c_{a}^{(N)}(t) +c_{b}^{(N)}(t)),\label{c-N-theta0}
\end{equation}
with $j=a,b$. If the two atoms are initially in the antisymmetric state, the concurrence $C^{(N)}_{-}$ retains its initial value $C^{(N)}_{-}(t)=C^{(N)}_{-}(0)=1$. When $\theta_{0}=(m+1/2)\pi$,  the concurrence $C_{\pm}^{(N)}(t)$ can be obtained as $C_{\pm }^{(N)}(t)=A_{\pm }e^{-2\gamma t}$ by solving Eqs.~(\ref{ca-N-td0}) and~(\ref{cb-N-td0}) under the initial condition, in which we introduce the modified coefficients $A_{\pm }=|(i+2)[\sqrt{4-2i}\sinh (2\gamma t\sqrt{-1-2i})\pm2\cosh (2\gamma t\sqrt{-1-2i})\mp i]/5|$.

By increasing the time delay to $\gamma t_{d}\sim1$, the non-Markovian retarded effect leads to some recovery oscillating peaks after experiencing an exponential decay within $\gamma t\in(0,\gamma t_{d})$, as shown in Figs.~\ref{CN-vs-tandtheta-3D}(c) and~\ref{CN-vs-tandtheta-3D}(d). In the long-time limit, the concurrence $C^{(N)}_{+}$ approaches to a steady-state value given in Eq.~(\ref{s-s value N 2}) when $\theta_{0}=(2m+1)\pi$, which implies the appearance of a subradiant state~\cite{Brewer96}. For the concurrence $C^{(N)}_{-}$, the condition for the appearance of the subradiant state is given by $\theta_{0}=m\pi$. However, the steady-state value of $C^{(N)}_{-}$ at $\theta_{0}=2m\pi$ and $\theta_{0}=(2m+1)\pi$ is determined by different expressions [see Eq.~(\ref{s-s value N 34})]. In Figs.~\ref{CN-vs-tandtheta-3D}(e) and~\ref{CN-vs-tandtheta-3D}(f), we find that the disentanglement dynamics between the two nested giant atoms exhibits the same exponential decay process as the previous coupling configurations with the time delay $\gamma t_{d}\rightarrow\infty$. Therefore, the influence of the coupling configuration on the disentanglement dynamics takes effect when the time delay is within an appropriate range, in which the two giant atoms can be re-excited by each other and each giant atom can re-absorb the photons radiated by itself.

For the nested coupling, we do not show the concurrences $C_{\pm}^{(N)}(t)$ as functions of the time delay $\gamma t_{d}$ and the evolution time $\gamma t$, since $C_{\pm}^{(N)}(t)$ exhibit similar characteristics with $C_{\pm}^{(S)}$ and $C_{\pm}^{(B)}$ when $\gamma t_{d}$ increases from $0$ to $\gamma t_{d}\sim1$.  Instead, we focus on the steady-state concurrence given in Eqs.~(\ref{s-s value N 2}) and~(\ref{s-s value N 34}), in which the steady-state concurrences $C_{\pm}^{(N)}(t\rightarrow\infty)$ are different from $C_{\pm}^{(S)}(t\rightarrow\infty)$ [Eqs.~(\ref{s-s value S 1}) and~(\ref{s-s value S 23})] and $C_{\pm}^{(B)}(t\rightarrow\infty)$ [Eqs.~(\ref{s-s value B 1}) and~(\ref{s-s value B 2})] at $\theta_{0}=(2m+1)\pi$. In Fig.~\ref{CN-vs-tandtheta-3D}, we display the $C^{(N)}_{\pm}(t\rightarrow\infty)$ given in Eqs.~(\ref{s-s value N 2}) and~(\ref{s-s value N 34}) as functions of $\gamma t_{d}$ when $\theta_{0}$ takes different values. Figure~\ref{CN-vs-tandtheta-3D} shows that the steady-state concurrences in Eqs.~(\ref{s-s value N 2}) and~(\ref{s-s value N 34}) satisfy the relation $C_{+}^{(N)}(\infty)_{\theta_{0}=(2m+1)\pi}>C_{-}^{(N)}(\infty)_{\theta_{0}=2m\pi}>C_{-}^{(N)}(\infty)_{\theta_{0}=(2m+1)\pi}$, where the value of the subscript $\theta_{0}$ is the condition for the appearance of the steady-state value. In particular, the steady-state value $C_{+}^{(N)}(\infty)_{\theta_{0}=(2m+1)\pi}$ is larger than those in the separate- and braided-coupling configurations, which indicates that the nested-coupling configuration is an optimal arrangement for achieving large steady-state entanglement in the two-giant-atom waveguide-QED system.

To clearly see the disentanglement dynamics between the two giant atoms for three different coupling configurations, we summarize in Table~\ref{table1} the expression of the concurrence in the considered cases under certain conditions. Based on Table~\ref{table1}, we know that the appearance of the steady-state entanglement depends on the coupling configurations, the phase shift, and the initial atomic state.
\begin{figure}[tbp]
\center\includegraphics[width=0.48\textwidth]{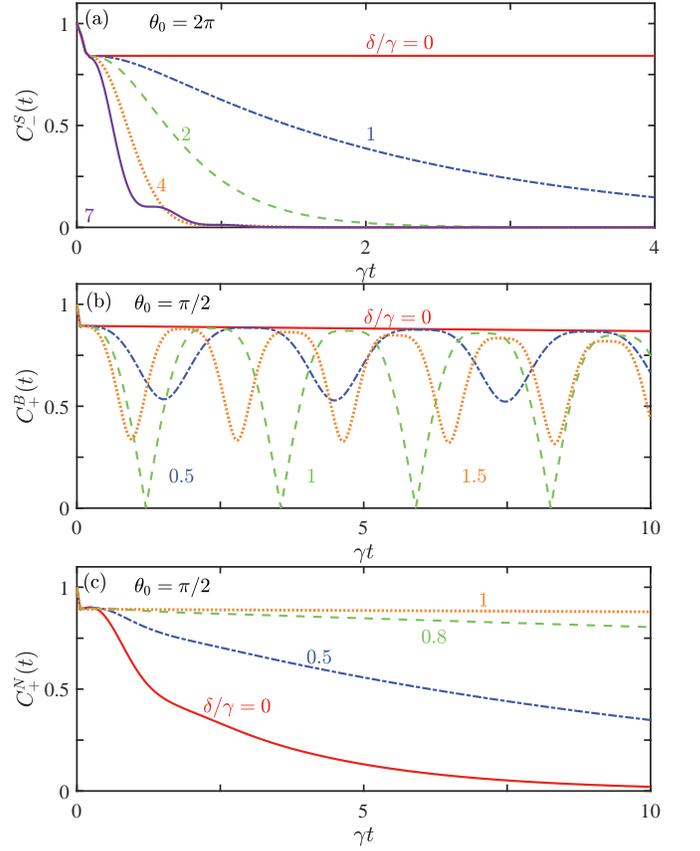}
\caption{(a)-(c) Concurrences $C^{(S)}_{-}(t)$, $C^{(B)}_{+}(t)$, and $C^{(N)}_{+}(t)$ versus the evolution time $\gamma t$ when $\delta/\gamma$ takes different values. In panel (a), we take $\theta_{0}=\pi$. In panels (b) and (c), we take $\theta_{0}=\pi/2$. In all panels, the time delay is $\gamma t_{d}=0.03$.}
\label{CBN-vs-tchgdelta}
\end{figure}

\section{Effect of the atomic detuning on the disentanglement dynamics of the two giant atoms}\label{Effect of the detuning}
In Sec.~\ref{Dynamics of disEn A}, we focus on the case where the two giant atoms have the same transition frequency.  Below, we consider that there exists a frequency detuning $\delta=\omega_{a}-\omega_{b}$ between the two giant atoms in three different coupling configurations. In Fig.~\ref{CBN-vs-tchgdelta}, we plot the concurrences $C_{-}^{(S)}(t)$,  $C_{+}^{(B)}(t)$, and  $C_{+}^{(N)}(t)$ as functions of $\gamma t$ when the detuning $\delta$ takes various values.  To better show the effect of the detuning on the disentanglement dynamics of the giant atoms, here we take a finite time delay. It can be seen from Fig.~\ref{CBN-vs-tchgdelta}(a) that the stationary value of the concurrence $C_{-}^{(S)}(t)$ is quickly destroyed with the increase of $\delta$ and that the $C_{-}^{(S)}(t)$ starts to exhibit oscillating decay. To explain the mechanism behind this feature, we take $\gamma t_{d}\rightarrow0$ and substitute $\theta_{0}=2\pi$ into Eqs.~(\ref{alpha-a-td0-S}) and~(\ref{alpha-b-td0-S}) to obtain
\begin{subequations}
\begin{align}
\dot{\alpha}_{+}^{(S)}(t)=&-i\delta \alpha _{-}^{(S)}(t) -4\gamma \alpha _{+}^{(S)}(t),\label{alpha-S-a-delta} \\
\dot{\alpha}_{-}^{(S)}(t)=&-i\delta \alpha _{+}^{(S)}(t).\label{alpha-S-b-delta}
\end{align}
\end{subequations}
From Eqs.~(\ref{alpha-S-a-delta}) and~(\ref{alpha-S-b-delta}), we find that the coupling between the states $|+\rangle$ and $|-\rangle$ is induced by the frequency detuning $\delta$. Therefore, the states $|+\rangle$ and $|-\rangle$  will exchange the population. For the symmetric state $|+\rangle$, there exists an additional dissipation channel with decay rate $4\gamma$ due to the coupling of the giant atoms to the continuous field modes in the waveguide. When $\delta/\gamma<4$, the concurrence $C_{-}^{(S)}(t)$ exhibits a monotonic decreasing behavior [see the green dashed and the blue dash-dotted curves in Fig.~\ref{CBN-vs-tchgdelta}(a)]. This is because the speed of the population exchange between the states $|+\rangle$ and $|-\rangle$ is smaller than the decay rate of the state $|+\rangle$, and then the population in the state $|+\rangle$ cannot be transferred to the state $|-\rangle$. However, when $\delta/\gamma>4$,  the population in the state $|+\rangle$ can come back to the state  $|-\rangle$, and then the concurrence $C_{-}^{(S)}(t)$ is characterized by an oscillating decay.

For the braided-coupling case, we can show that the amplitudes $\alpha _{\pm}^{(B)}(t)$ have the same time evolution with $\alpha _{\pm}^{(S)}(t)$ given in Eqs.~(\ref{alpha-S-a-delta}) and~(\ref{alpha-S-b-delta}) by substituting  $\theta_{0}=2\pi$ into Eq.~(\ref{S-ANS-Bpm}). It was shown in Ref.~\cite{Kockum18} that for $\theta_{0}=\pi/2$, there exist vanished individual decays for the two braided giant atoms and a nonzero exchanging interaction (called the decoherence-free interaction~\cite{Kockum18}) between them. Therefore, we show the concurrence $C_{+}^{(B)}(t)$ in Fig.~\ref{CBN-vs-tchgdelta}(b) as a function of $\gamma t$ at given values of $\delta/\gamma$ when $\theta_{0}=\pi/2$. The equations of motion for the amplitudes $\alpha _{\pm}^{(B)}(t)$ at $\theta_{0}=\pi/2$ and $\gamma t_{d}\rightarrow0$ become
\begin{subequations}
\begin{align}
\dot{\alpha}_{+}^{(B)}(t) =&-i\delta \alpha _{-}^{(B)}(t)-i\gamma \alpha_{+}^{(B)}(t), \label{alpha-B-a-delta} \\
\dot{\alpha}_{-}^{(B)}(t) =&-i\delta \alpha _{+}^{(B)}(t)+i\gamma \alpha_{-}^{(B)}(t). \label{alpha-B-b-delta}
\end{align}
\end{subequations}
By solving Eqs.~(\ref{alpha-B-a-delta}) and~(\ref{alpha-B-b-delta}) under the initial condition [$\alpha _{+}^{(B)}(0)=1$ or $\alpha _{-}^{(B)}(0)=1$], the concurrence can be obtained as
\begin{equation}
C_{\pm }^{(B)}(t) =\frac{\sqrt{[\gamma ^{2}+\delta ^{2}\cos(2\Omega t) ]^{2}+\delta ^{2}\Omega ^{2}\sin ^{2}(2\Omega t) }}{\Omega ^{2}},\label{CB-delta}
\end{equation}
where we introduce $\Omega =\sqrt{\gamma ^{2}+\delta ^{2}}$. According to Eqs.~(\ref{alpha-B-a-delta}),~(\ref{alpha-B-a-delta}), and~(\ref{CB-delta}), it can be found that the concurrences $C_{\pm}^{B}(t)$ can preserve their initial value when $\delta=0$, which is consistent with our previous analysis. However, when $\delta\neq0$, the initially occupied state $|+\rangle$ ($|-\rangle$) exchanges energy with the state $|-\rangle$ ($|+\rangle$). Since there is no additional dissipation channel for the states $|\pm\rangle$, and hence the concurrences $C_{\pm}^{B}(t)$ undergo periodic oscillations with a period $\pi /\sqrt{\gamma ^{2}+\delta ^{2}}$. Figure~\ref{CBN-vs-tchgdelta}(b) shows that the oscillation period of the concurrence $C_{+}^{B}(t)$ increases with the increase of $\delta$. In particular, the energy exchange between the states $|+\rangle$ ($|-\rangle$) and $|-\rangle$ ($|+\rangle$) can reach the maximum when $\delta/\gamma=1$.

In the case of nested giant atoms, the amplitudes $\alpha_{\pm}^{(N)}(t)$ also have the same time evolution with $\alpha_{\pm}^{(S)}(t)$ described by Eqs.~(\ref{alpha-S-a-delta}) and~(\ref{alpha-S-b-delta}) when we take $\theta_{0}=2\pi$ in Eqs.~(\ref{amplitude-ca-N}) and~(\ref{amplitude-cb-N}). Figure~\ref{CBN-vs-tchgdelta}(c) shows the concurrence $C_{+}^{N}(t)$ as a function of $\gamma t$ when $\delta$ takes different values and $\theta_{0}=\pi/2$. Considering the time delay $\gamma t_{d}\rightarrow0$, Eqs.~(\ref{amplitude-ca-N}) and~(\ref{amplitude-cb-N}) are reduced to
\begin{subequations}
\begin{align}
\dot{\tilde{c}}_{a}^{(N)}(t) =&-[\gamma +i(\delta-\gamma)]\tilde{c}_{a}^{(N)}(t)-i\gamma \tilde{c}_{b}^{(N)}(t)+\gamma \tilde{c}_{b}^{(N)}(t),\label{ca-N-delta} \\
\dot{\tilde{c}}_{b}^{(N)}(t) =&-\left[ \gamma -i(\delta -\gamma )\right]\tilde{c}_{b}^{(N)}(t)-i\gamma \tilde{c}_{a}^{(N)}(t)+\gamma \tilde{c}_{a}^{(N)}(t).\label{cb-N-delta}
\end{align}
\end{subequations}
If we further consider the condition $\delta/\gamma=1$, then Eqs.~(\ref{ca-N-delta}) and~(\ref{cb-N-delta}) become
\begin{subequations}
\begin{align}
\dot{\alpha}_{+}^{(N)}(t) =&-i\delta\alpha_{+}^{(N)}(t), \label{alpha-p-N-delta}\\
\dot{\alpha}_{-}^{(N)}(t)=&-2\delta\alpha_{-}^{(N)}(t)-i\delta \alpha_{-}^{(N)}(t),\label{alpha-m-N-delta}
\end{align}
\end{subequations}
 where we introduce the symmetric and antisymmetric amplitudes $\alpha _{\pm }^{(N)}(t)=[\tilde{c}_{a}^{(N)}(t) \pm \tilde{c}_{b}^{(N)}(t)]/\sqrt{2}$ for the two nested giant atoms. Equations~(\ref{alpha-p-N-delta}) and~(\ref{alpha-m-N-delta}) indicate that  the amplitudes $\alpha _{\pm }^{(N)}(t)$ are decoupled from each other, and hence the concurrence $C_{+}^{N}(t)$ can still preserve its initial value even when there exists the frequency detuning between the two nested giant atoms [see the orange dashed curve in Fig.~(\ref{CBN-vs-tchgdelta})], while this feature does not exist for small atoms. When the nested giant atoms are initially in the antisymmetric state $|-\rangle$, the concurrence $C_{-}^{N}(t)$ will exhibit exponentially decays with time at a rate $2\delta$. These results indicate that the frequency detuning of the giant atoms can also play an important role in the manipulation of quantum entanglement between the two giant atoms.
\begin{figure}[tbp]
\center\includegraphics[width=0.48\textwidth]{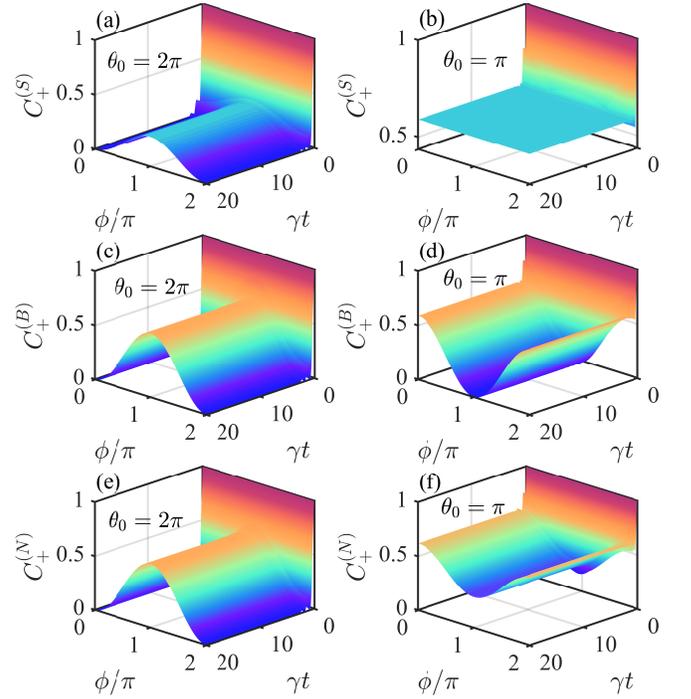}
\caption{Concurrences (a,b) $C_{+}^{(S)}$, (c,d) $C_{+}^{(B)}$, and (e,f) $C_{+}^{(N)}$ as functions of the evolution time $\gamma t$ and the phase $\phi$ when the two giant atoms are initially in the state $|\psi\rangle _{+}=(|e\rangle _{a}|g\rangle _{b}+ e^{i\phi }|g\rangle _{a}|e\rangle _{b})/\sqrt{2}$. The left and right columns correspond to the phase $\theta_{0}=2\pi$ and $\pi$, respectively. In all panels, we take $\gamma t_{d}=0.3$.}
\label{CSBN-vs-tandphi-sup}
\end{figure}

\section{Effect of the initial-state superposition phase $\phi$ on the disentanglement dynamics of the two giant atoms in the single-excitation space}\label{Effect of the phase}
In Secs.~\ref{Dynamics of disEn A} and~\ref{Effect of the detuning}, we focus on the disentanglement dynamics of the two giant atoms starting in the symmetric and antisymmetric states. We next study the disentanglement dynamics of the two giant atoms in a general pure entangled state $|\psi\rangle _{+}=(|e\rangle _{a}|g\rangle _{b}+ e^{i\phi }|g\rangle _{a}|e\rangle _{b})/\sqrt{2}$ in the single-excitation space. As shown in Figs.~\ref{CSBN-vs-tandphi-sup}(a) and~\ref{CSBN-vs-tandphi-sup}(b), we plot the concurrence $C^{(S)}_{+}$ as a function of $\gamma t$ and $\phi$ when $\theta_{0}=2\pi$ and $\pi$, respectively. From Fig.~\ref{CS-vs-tandtheta-3D}(a), we can see that the steady-state entanglement monotonically increases with $\phi\in[0,\pi]$ and decreases with $\phi\in[\pi,2\pi]$ when $\theta_{0}=2\pi$. However, when we take $\theta_{0}=\pi$, the steady-state entanglement becomes $\phi$-independent. By using the final-value theorem, the dependence of the steady-state entanglement on the phase $\phi$ at $\theta_{0}=2m\pi$ can be obtained as
\begin{equation}
C_{+}^{(S)}(t\rightarrow \infty )=\frac{1-\cos \phi }{2( 1+3\gamma t_{d}) ^{2}},\ \ \ \theta _{0}= 2m\pi.\label{S-C-phi}
\end{equation}
Equation~(\ref{S-C-phi}) indicates that when $\phi=\pi$, the stationary value of the concurrence $C_{+}^{(S)}(t)$ can reach the maximal value, which is consistent with the numerical results in Fig.~\ref{CSBN-vs-tandphi-sup}(a).

To see the effect of the phase $\phi$ on the disentanglement dynamics of two braided giant atoms, in Figs.~\ref{CSBN-vs-tandphi-sup}(c) and~\ref{CSBN-vs-tandphi-sup}(d) we plot the concurrence $C_{+}^{(B)}$ as a function of $\gamma t$ and $\phi$ when $\theta_{0}=2\pi$ and $\pi$, respectively. Different from the separate case, here we find that the steady-state entanglement at $\theta_{0}=2\pi$ and $\pi$ are both $\phi$-dependent. By resorting to the final-value theorem, we obtain the $\phi$-dependent steady-state entanglement as
\begin{subequations}
\begin{align}
C_{+}^{(B)}(t &\rightarrow \infty )=\frac{1+\cos \phi }{2(1+\gamma t_{d}) ^{2}},\ \ \ \theta _{0}=(2m+1)\pi , \label{C-B-phi2mpus1pi}  \\
C_{+}^{(B)}(t &\rightarrow \infty )=\frac{1-\cos \phi }{2(1+\gamma t_{d}) ^{2}},\ \ \ \theta _{0}= 2m\pi. \label{C-B-phi2mpi}
\end{align}
\end{subequations}
Equation~(\ref{C-B-phi2mpus1pi}) indicates that the concurrence  $C_{+}^{(B)}$ can reach the maximal stationary value given by Eq.~(\ref{s-s value B 1}) when $\phi=0$ and $2\pi$, which corresponds to the initial state $|+\rangle$ of the two giant atoms. However, according to Eq.~(\ref{C-B-phi2mpi}), the maximal stationary value of $C_{+}^{(B)}$ is obtained at $\phi=\pi$, which corresponds to the result given by Eq.~(\ref{s-s value B 2}), where the two giant atoms are initially in the state $|-\rangle$.

Figures~\ref{CSBN-vs-tandphi-sup}(e) and~\ref{CSBN-vs-tandphi-sup}(f) show the concurrence $C^{(N)}_{+}$ versus $\gamma t$ and $\phi$ when $\theta_{0}=2\pi$ and $\pi$, respectively. By applying the final-value theorem, the $\phi$-dependent steady-state entanglement are given by
\begin{subequations}
\begin{align}
C_{+}^{(N)}(t &\rightarrow\infty )=\frac{f(\phi ,\gamma t_{d})}{(1+4\gamma t_{d}+2\gamma ^{2}t_{d}^{2}) ^{2}},\hspace{0.2cm}\theta _{0}=(2m+1) \pi,\label{C-N-phi2mpus1pi}\\
C_{+}^{(N)}(t &\rightarrow\infty )=\frac{1-\cos \phi }{2(1+\gamma t_{d}) ^{2}},\hspace{1.3cm}\theta _{0}=2m\pi,\label{C-N-phi2mpi}
\end{align}
\end{subequations}
where the function $f(\phi ,\gamma t_{d})=|[\gamma t_{d}( e^{i\phi }+1) +1][\gamma t_{d}( 3e^{-i\phi }+1) +e^{-i\phi }]|$ is introduced. We find that the steady-state values of $C_{+}^{(N)}$ have an identical $\phi$-dependent relation with that of $C_{+}^{(B)}$ when $\theta_{0}=2\pi$ [see Eqs.~(\ref{C-B-phi2mpi}) and~(\ref{C-N-phi2mpi})]. In particular,  when $\phi=\theta_{0}=\pi$, the minimal stationary value of $C^{(N)}_{+}$ is nonzero, which is different from the case of the braided giant atoms. By substituting $\phi=\pi$ into Eq.~(\ref{C-N-phi2mpus1pi}), the minimal stationary entanglement is equal to $C_{-}^{(N)}(\infty )_{\theta _{0}=\pi }$ when the nested giant atoms are initially in the state $|-\rangle$ [see Eq.~(\ref{s-s value N 34})]. According to the $\phi$-dependent steady-state entanglement of three different coupling configurations, we know that a large steady-state entanglement can be achieved when the two giant atoms are initially prepared in symmetric or antisymmetric states.
\begin{figure}[tbp]
\center\includegraphics[width=0.48\textwidth]{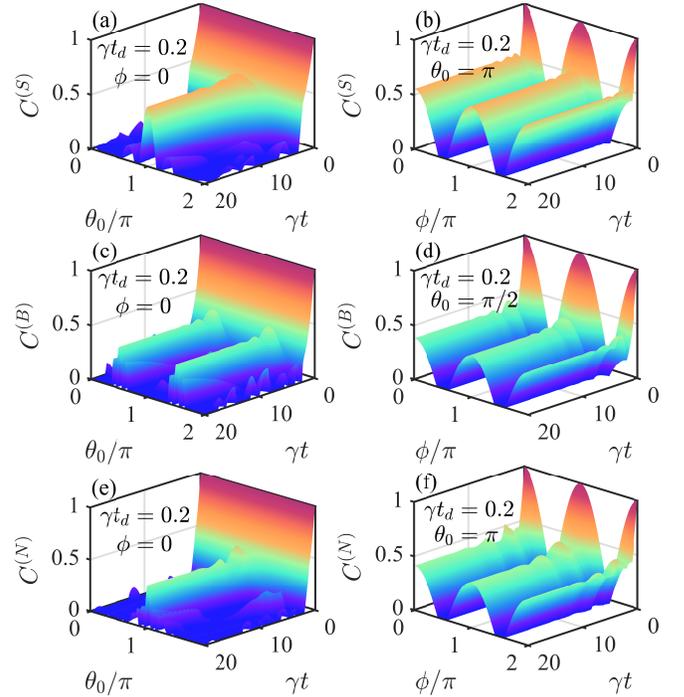}
\caption{(a), (c), and (e) Concurrences $C^{(S)}$,  $C^{(B)}$, and  $C^{(N)}$ as functions of the evolution time $\gamma t$ and the phase shift $\theta_{0}$ when $\phi=0$. (b), (d), and (f) Concurrences $C^{(S)}$,  $C^{(B)}$, and  $C^{(N)}$ as functions of $\gamma t$ and $\phi$. In panels (b), (d), and (f), the phase shifts are $\theta_{0}=\pi$, $\pi/2$, and $\pi$, respectively. The other parameter is $\gamma t_{d}=0.2$.}
\label{CSBN_ee_plus_gg}
\end{figure}

\section{Disentanglement dynamics of the two giant atoms in the initial state $(|g\rangle _{a}|g\rangle _{b}+e^{i\phi}|e\rangle _{a}|e\rangle _{b}) /\sqrt{2}$}\label{Double-extication state}
In previous sections, we focused on the single-excitation subspace of the system and obtained the evolution of the two giant atoms by solving the time-delayed equations of motion for the probability amplitudes. Here we present  numerical results of the disentanglement dynamics of the two giant atoms when they are initially in the state $|\tilde{\psi}\rangle _{+}=(|g\rangle _{a}|g\rangle _{b}+e^{i\phi}|e\rangle _{a}|e\rangle _{b}) /\sqrt{2}$. This is achieved by numerically solving the time-delayed quantum master equation of two giant atoms for three different coupling configurations. We would like to point out that the time-delayed quantum master equation for the two giant atoms can be obtained by applying the method used in Refs.~\cite{Zhu21,Yin22}. In the left column in Fig.~\ref{CSBN_ee_plus_gg}, we plot the concurrences $C^{(S)}$,  $C^{(B)}$, and  $C^{(N)}$ as functions of $\theta_{0}$ and $\gamma t$ when $\gamma t_{d}=0.2$. For the steady-state entanglement of the giant atoms in these three coupling configurations, we first take $\phi=0$ in Figs.~\ref{CSBN_ee_plus_gg}(a),~\ref{CSBN_ee_plus_gg}(b), and~\ref{CSBN_ee_plus_gg}(c), corresponding to the Bell state $|\Phi\rangle_{+}=(|g\rangle _{a}|g\rangle _{b}+|e\rangle _{a}|e\rangle _{b}) /\sqrt{2}$. It can be seen that both the concurrences $C^{(S)}$ and $C^{(N)}$ can reach a stationary value at $\theta_{0}=(2m+1)\pi$ for an integer $m$ in the long-time limit. However, for the two braided giant atoms, the steady-state value of  $C^{(B)}$ appears at $\theta_{0}=(m+1/2)\pi$. In the long-time limit, the steady-state entanglement between the giant atoms for these three coupling configurations satisfies the relation $C^{(S)}(\infty )_{\theta _{0}=(2m+1)\pi }>C^{(N)}(\infty )_{\theta_{0}=(2m+1)\pi }>C^{(B)}(\infty )_{\theta _{0}=(m+1/2)\pi }$.

To see the effect of the phase $\phi$ in the state $|\tilde{\psi}\rangle _{+}$ on the steady-state entanglement, we show the concurrences $C^{(S)}$,  $C^{(B)}$, and  $C^{(N)}$ as functions of $\phi$ and $\gamma t$ in the right column in Fig.~\ref{CSBN_ee_plus_gg}, when the phase shift $\theta_{0}=\pi$ in Figs.~\ref{CSBN_ee_plus_gg}(b) and~\ref{CSBN_ee_plus_gg}(f) and $\pi/2$ in Fig.~\ref{CSBN_ee_plus_gg}(d), respectively. In these phase shifts, the steady-state entanglement exists in the long-time limit. From Figs.~\ref{CSBN_ee_plus_gg}(b), ~\ref{CSBN_ee_plus_gg}(d), and~\ref{CSBN_ee_plus_gg}(f), we see that the two giant atoms have similar entanglement evolutions but different stationary values for the three coupling configurations. Meanwhile, these values approach the maximum at $\phi=0$ and $\pi$, which corresponds to the giant atoms starting in the Bell states $|\Phi\rangle_{\pm}=(|g\rangle _{a}|g\rangle _{b}\pm|e\rangle _{a}|e\rangle _{b}) /\sqrt{2}$. However, the stationary value of  $C^{(S)}$,  $C^{(B)}$, and  $C^{(N)}$ vanish in the long-time limit when $\phi$ is near to $(m+1/2)\pi$, which is different from the case of the single-excitation state, as shown in Fig.~\ref{CSBN-vs-tandphi-sup}.

\section{Discussion and conclusion}\label{Discussion and conclusion}
We present some discussions of the experimental implementation of the three kinds of double-giant-atom waveguide-QED systems. It was reported that the giant atoms can be realized in experiments by coupling the superconducting qubits to the SAWs~\cite{Delsing14,Leek17,Cleland19,Delsing19,Cleland20,Delsing20} or microwave waveguides~\cite{Oliver20,Wilson21}.  Therefore, these three setups can be used to implement the present scheme. Concretely, we can utilize two frequency-tunable Xmon qubits to couple with a coplanar microwave waveguide~\cite{Kockum14PRA,Oliver20,Wilson21}. The accumulated phase shift between two neighboring coupling points can be adjusted by tuning the qubit frequencies. To include the non-Markovian effect in this system, the distances between two neighboring coupling points need to be of order of $d\approx10$ m to satisfy the time delay $\gamma t_{d}\sim1$~\cite{Oliver20}. In addition, we can also use two transmon qubits to couple with a SAWs transmission waveguide through multiple interdigital transducers~\cite{Delsing19,Delsing20}.  Due to the slow propagating speed of the SAWs, the time delay between two neighboring coupling points of the giant atoms becomes remarkable. In Ref.~\cite{Delsing19}, the time delay was realized to reach $\gamma t_{d}\approx14$, which is well in the non-Markovian regime. All these advances indicate that the three kinds of systems in our work is experimentally accessible with current and near-future conditions.

In conclusion, we studied the disentanglement dynamics of two coupled to a waveguide with three different coupling configurations. We considered the influence of the non-Markovian retarded effect on the disentanglement dynamics between two giant atoms in this double-giant-atom waveguide-QED system. Concretely, we considered three coupling configurations: the separate, braided, and nested couplings. It was shown that the evolution of the entanglement can exhibit oscillating decay or steady-state value by adjusting the accumulated phase shift, the initial atomic state, and the coupling configurations. The appearance of the steady-state entanglement indicates the existence of the subradiant state.  We obtained the expressions of the time-delay-dependent steady-state concurrence between the two giant atoms, which shows that the increase of the time delay will degrade the value of the steady-state concurrence. For the non-resonance case, the frequency detuning plays a crucial role in the manipulation of the disentanglement dynamics, which can exhibit different characteristics from that in small atoms due to quantum interference effect induced by the multipled coupling points of giant atoms. This work will pave the way for generating steady-state entanglement between giant atoms based on the giant-atom waveguide-QED systems, which can be used as a significant resource to apply in modern quantum technologies, such as quantum computation, communication, and metrology~\cite{Duarte21}.

\begin{acknowledgments}
The authors thank Y. T. Zhu for helpful discussions. J.-Q.L. was supported, in part, by the National Natural Science Foundation of China (Grants No.~12175061, No.~11822501, No.~11774087, No.~12247105, and No.~11935006), the Science and Technology Innovation Program of Hunan Province (Grants No.~2021RC4029 and No.~2020RC4047), and the Hunan Science and Technology Plan Project (Grant No.~2017XK2018).
\end{acknowledgments}

\end{document}